\documentclass[10pt, conference, compsocconf]{IEEEtran}
\def\IEEEbibitemsep{0pt plus .5pt}

\usepackage{balance}
\usepackage{array,enumerate}
\usepackage{multirow}
\usepackage{graphicx}
\usepackage[linesnumbered,ruled,vlined]{algorithm2e}
\usepackage[footnotesize]{caption}
\usepackage{float}
\SetArgSty{textup}
\usepackage{amssymb}
\usepackage{amsmath}
\usepackage{url}
\usepackage{threeparttable}
\usepackage{scrextend}
\usepackage{bm}
\usepackage{tikz}
\usepackage{algpseudocode}
\usepackage{ragged2e}
\usepackage{courier}
\usepackage{times}

\usepackage{array}
\newcolumntype{L}[1]{>{\raggedright\let\newline\\\arraybackslash\hspace{0pt}}m{#1}}
\newcolumntype{C}[1]{>{\centering\let\newline\\\arraybackslash\hspace{0pt}}m{#1}}
\newcolumntype{R}[1]{>{\raggedleft\let\newline\\\arraybackslash\hspace{0pt}}m{#1}}

\SetCommentSty{mycommfont}

\newcommand{\subcaption}[1]{\centerline{{\scriptsize
  #1}}\vspace{10pt}}
\newlength{\minipagewidth}
\newlength{\figurewidthFour}

\begin{document}

\title{GraphH: High Performance Big Graph Analytics in Small Clusters}


\author{
    \IEEEauthorblockN{Peng\ Sun,
    Yonggang\ Wen,
    Ta\ Nguyen Binh Duong
    and Xiaokui Xiao
    }
    \IEEEauthorblockA{School of Computer Science and Engineering, Nanyang Technological University, Singapore
    \IEEEauthorblockA{\{sunp0003, ygwen, donta, xkxiao\}@ntu.edu.sg}}
}

\maketitle

\thispagestyle{plain}
\pagestyle{plain}

\begin{abstract}

It is common for real-world applications to analyze big graphs using distributed graph processing systems.  Popular in-memory systems require an enormous amount of resources to handle big graphs. 
While several out-of-core approaches have been proposed for processing big graphs on disk, the high disk I/O overhead could significantly reduce performance. In this paper, we propose GraphH to enable high-performance big graph analytics in small clusters. Specifically, we design a two-stage graph partition scheme to evenly divide the input graph into partitions, and propose a GAB (Gather-Apply-Broadcast) computation model to make each worker process a partition in memory at a time. We use an edge cache mechanism to reduce the disk I/O overhead, and design a hybrid strategy to improve the communication performance. GraphH can efficiently process big graphs in small clusters or even a single commodity server. Extensive evaluations have shown that GraphH could be up to 7.8x faster compared to popular in-memory systems, such as Pregel+ and PowerGraph when processing generic graphs, and more than 100x faster than recently proposed out-of-core systems, such as GraphD and Chaos when processing big graphs.

\end{abstract} 


\section{Introduction}\label{sec: introduction}

Many distributed graph processing systems have been proposed to tackle general graph analytics  in memory. 
They usually follow the ``think like a vertex''  philosophy, and abstract graph computation as vertex-centric programs.
More specifically, Pregel \cite{malewicz2010pregel}, Giraph \cite{ching2015one}, Pregel+ \cite{yan2014pregelplus},  GPS \cite{salihoglu2013gps}, MOCGraph \cite{zhou2014mocgraph} and HuSky \cite{yang2016husky} adopt the Pregel computation model: they assign the input graph's vertices to multiple machines, and provide interaction between vertices by message passing along out-edges. 
PowerGraph \cite{gonzalez2012powergraph}, PowerLyra \cite{chen2015powerlyra}, GraphX \cite{gonzalez2014graphx} and LFGraph \cite{hoque2013lfgraph} use the GAS (Gather-Apply-Scatter)  model: they split a vertex into multiple replicas, and parallelize the computation for a single vertex in different machines. 
Many benchmarking results have shown that these  systems could offer better performance than general-purpose systems like Hadoop and Spark \cite{bench1}, \cite{bench2}, \cite{hu2014toward}.

Aforementioned in-memory approaches require powerful computation resources to process and analyze big graphs\footnote{A big graph usually contains billions of vertices or hundreds of billions of edges.},
neglecting the need of average users who cannot  afford  a large cluster. 
During the computation, these systems need to store the entire input graph and all network-transmitted messages in memory.
This strategy is appropriate when processing generic graphs with a few billion edges, such as  {Twitter-2010} and {UK-2007} as shown in Table \ref{Tab: Datasets}.
Unfortunately, it is common for real-world applications to process and analyze big graphs like {UK-2014 and {EU-2015}, which are orders of magnitude larger than Twitter-2010 for example.
In this case, the input graph and intermediate messages can easily exceed the memory limit of a small-scale cluster, leading to significant performance degradation or even program crashes. 
We evaluated the memory requirement for running PageRank on {UK-2007} with five in-memory graph processing systems in a 9-node cluster. As shown in Figure \ref{Fig: MemUsage_and_Time} (a), Giraph, GraphX, PowerGraph, PowerLyra and Pregel+ need 795GB, 685GB,  357GB, 511GB and  281GB memory,  indicating 8.5x, 7.3x, 3.8x, 5.5x and 2.9x memory explosions with respect to the input graph's size.
To process big graphs like EU-2015, these in-memory approaches require a large cluster with at least $5$TB memory, approximately.


\renewcommand\arraystretch{1}
\begin{table} 
\centering
\resizebox{0.49\textwidth}{!}{
\begin{threeparttable}{}
\caption{Benchmark Graph Datasets.}
\begin{tabular}{|l|r|r|r|r|r|r|}
\hline
\textbf{Graphs}  & \begin{tabular}[c]{@{}c@{}} \textbf{Vertex} \\ \textbf{Num} \end{tabular}   & \begin{tabular}[c]{@{}c@{}} \textbf{Edge} \\ \textbf{Num} \end{tabular}   & \begin{tabular}[c]{@{}c@{}} \textbf{Avg} \\ \textbf{Deg} \end{tabular}  & \begin{tabular}[c]{@{}c@{}} \textbf{Max} \\ \textbf{Indeg} \end{tabular} & \begin{tabular}[c]{@{}c@{}} \textbf{Max} \\ \textbf{Outdeg} \end{tabular} &  \begin{tabular}[c]{@{}c@{}} \textbf{Size} \\ \textbf{(CSV)} \end{tabular}  \\ \hline
Twitter-2010      & 42M   & 1.5B  & 35.3 &0.7M &770K & 25GB      \\
UK-2007      & 134M   & 5.5B  & 41.2 &6.3M &22.4K & 93GB      \\ 
UK-2014      & 788M   & 47.6B & 60.4 &8.6M &16.3K  &0.9TB     \\
EU-2015      & 1.1B & 91.8B & 85.7 &20M &35.3K & 1.7TB \\ \hline
\end{tabular}
\label{Tab: Datasets}
\end{threeparttable}
}
\end{table}

\setlength{\minipagewidth}{0.237\textwidth}
\setlength{\figurewidthFour}{\minipagewidth}
\begin{figure} 
    \centering
    \begin{minipage}[t]{\minipagewidth}
    \begin{center}
    \includegraphics[width=\figurewidthFour]{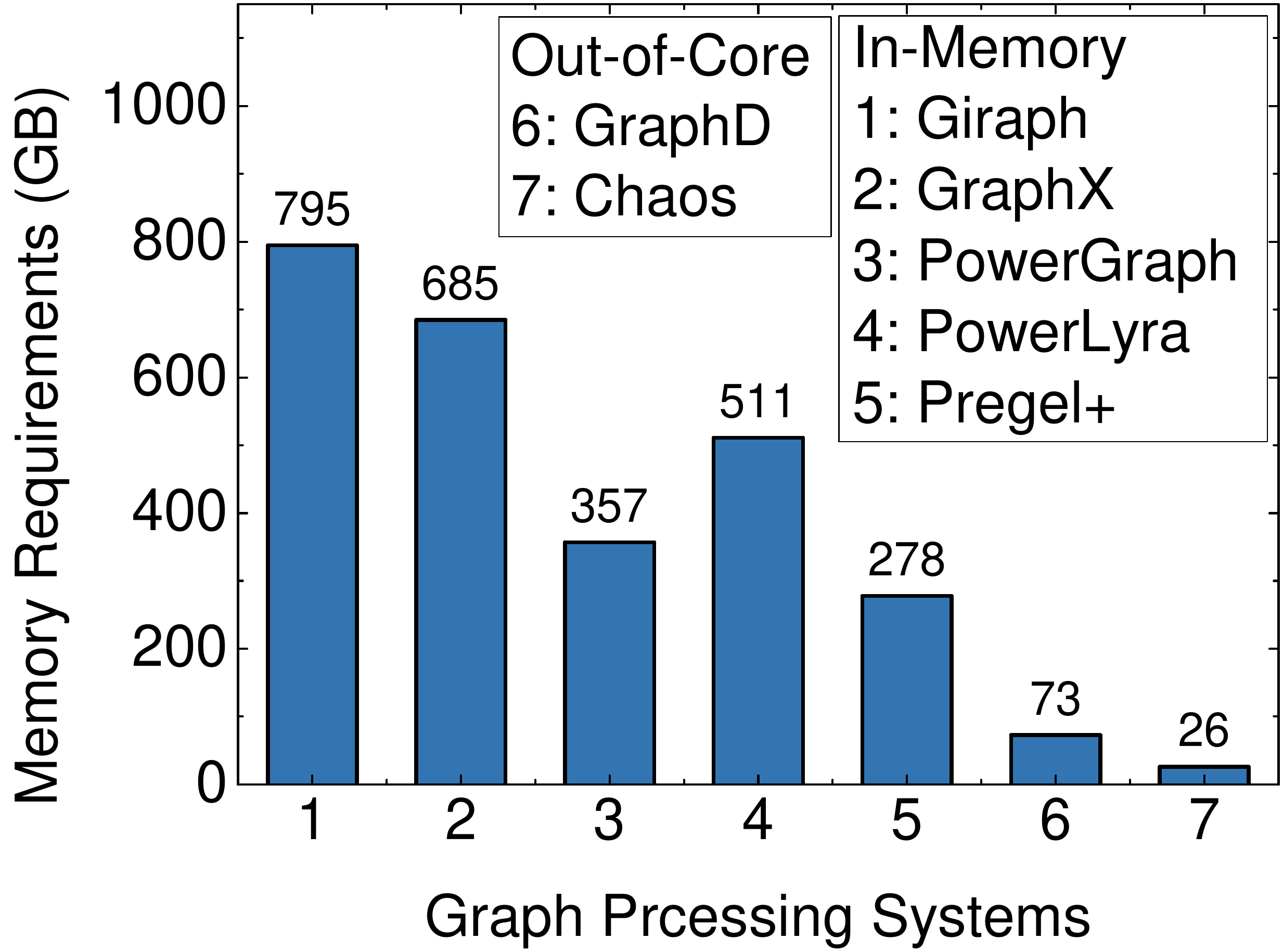}
    \subcaption{(a) Memory Requirements (UK-2007)}
    \end{center}
    \end{minipage}
    \centering
    \begin{minipage}[t]{\minipagewidth}
    \begin{center}
    \includegraphics[width=\figurewidthFour]{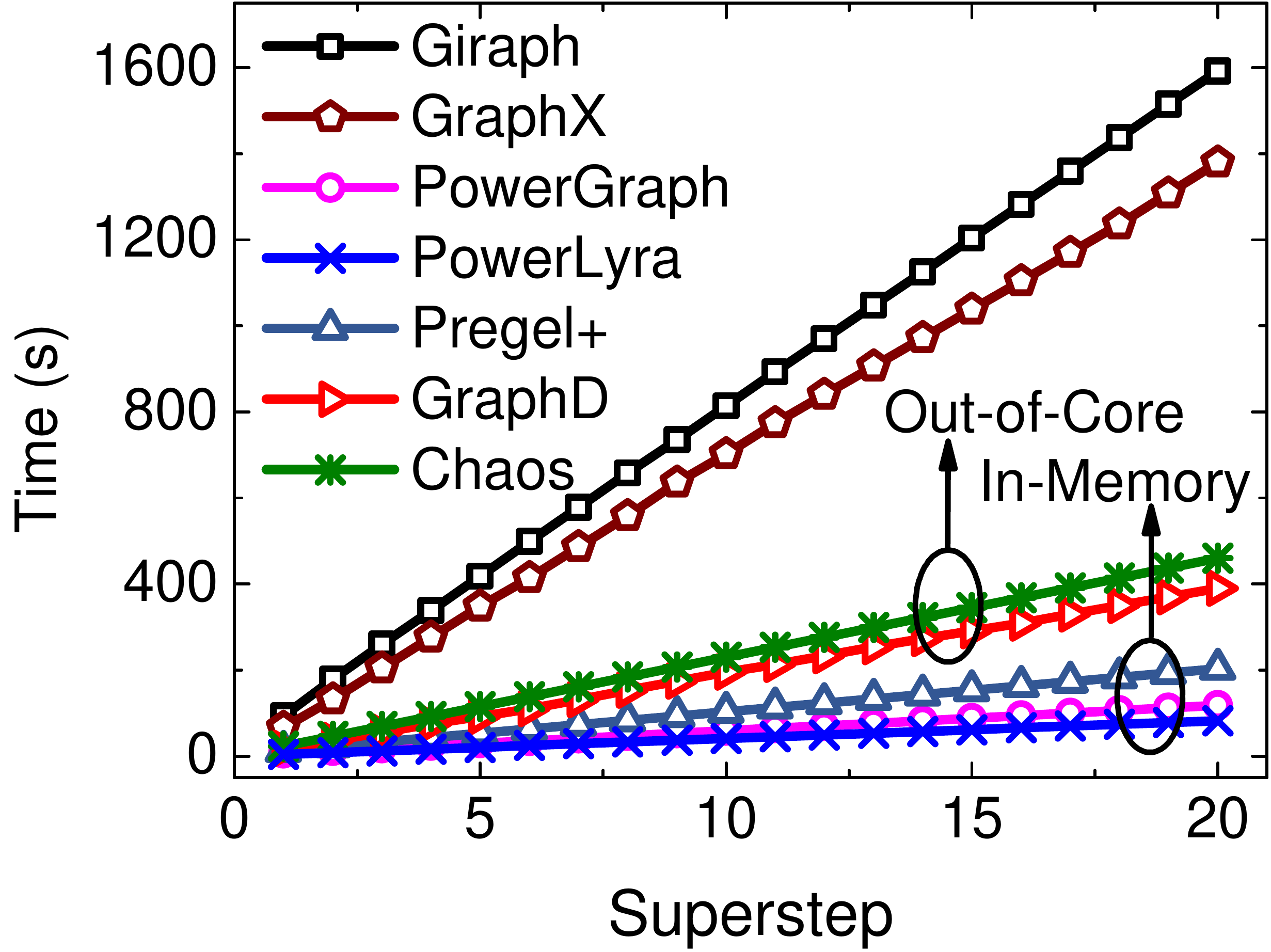}
    \subcaption{(b)  Execution Time (UK-2007)}
    \end{center}
    \end{minipage}
    \centering
    \caption{Memory requirements and execution time for running PageRank on {UK-2007} with various distributed graph processing systems. The testbed has 9 servers, and  each server contains 12x2.0GHz cores (two Intel Xeon E5-2620 CPUs), 128GB memory, 4x4TB HDDs (RAID5) and 10Gbps Ethernet. We use Giraph-1.1, GraphX-2.0, PowerGraph-2.2  and the newest version of PowerLyra (hybrid-ginger mode), Pregel+, GraphD and Chaos.}
\label{Fig: MemUsage_and_Time}
\end{figure}

\renewcommand\arraystretch{1.1}
\begin{table*} 
\centering
\resizebox{1\textwidth}{!}{
\begin{threeparttable}{}
\caption{A Comparison of Different Distributed Graph Processing Systems.}
\label{my-label}
\begin{tabular}{|l|l|l|l|}
\hline
 & \begin{tabular}[c]{@{}c@{}} \textbf{In-Memory} \end{tabular} & \begin{tabular}[c]{@{}c@{}} \textbf{Out-of-Core} \end{tabular} & \begin{tabular}[c]{@{}c@{}} \textbf{Hybrid} \end{tabular} \\ \hline \hline
\begin{tabular}[l]{@{}l@{}} \textbf{Distributed Graph} \\ \textbf{Processing System} \end{tabular} & \begin{tabular}[l]{@{}l@{}} Pregel, Pregel+, Giraph, GPS, MOCGraph, HuSky \\ GraphX, LFGraph, PowerGraph, PowerLyra   \end{tabular} & \begin{tabular}[l]{@{}l@{}} GraphD, Chaos \end{tabular} &  \begin{tabular}[l]{@{}l@{}} GraphH (\emph{Our Approach})  \end{tabular} \\ \hline
\textbf{In Memory Data} & \begin{tabular}[l]{@{}l@{}}All Vertex States, Edges $\&$  Messages \end{tabular} &  \begin{tabular}[l]{@{}l@{}} (Part) Vertex States \end{tabular}   &  \begin{tabular}[l]{@{}l@{}} All Vertex States  $\&$ Messages, Cached Edges   \end{tabular} \\ \hline
\textbf{Required Platform} & Large-Scale  Clusters or Supercomputers & Small Commodity Clusters  & Small Commodity  Clusters  \\ \hline
\textbf{Performance} &  High (no disk I/O during computation)  &  Low (frequent disk I/O)  &  High (reduce disk I/O by cache)  \\ \hline
\end{tabular}
\label{Tab: Compare}
\end{threeparttable}
}
\end{table*}

Researchers have proposed several out-of-core systems to enable big graph analytics with limited memory.
Specifically, single-node systems, such as GraphChi \cite{kyrola2012graphchi}, VENUS \cite{cheng2015venus}, X-Stream \cite{roy2013x} and GridGraph \cite{zhu2015gridgraph}, can process big graphs from secondary storage on a single server. 
GraphD \cite{yan2016efficient} and Chaos \cite{roy2015chaos} could scale out-of-core graph processing to multiple servers in a cluster. 
These out-of-core systems typically maintain vertices in memory, and manage edges/messages on disks to reduce memory footprint.
As shown in Figure \ref{Fig: MemUsage_and_Time} (a), GraphD and Chaos use 73GB and 26GB memory  to run PageRank on {UK-2007}, respectively. Both systems can process UK-2014 and EU-2015 in a 9-node cluster.

However, existing out-of-core systems could incur a huge amount of disk accesses,  resulting in significant performance reduction. 
As shown in Figure \ref{Fig: MemUsage_and_Time} (a), while each server has 128GB memory, GraphD and Chaos only use 8GB and 3GB memory per server respectively, and cannot efficiently leverage idle memory to reduce disk I/O overhead. 
Thus, out-core systems usually have much lower performance than most in-memory systems. 
As shown in Figure \ref{Fig: MemUsage_and_Time} (b), PowerGraph, PowerLyra and Prgel+  outperform GraphD by 3.3x, 4.8x, and 1.9x, and outperform Chaos by 3.8x, 5.6x and 2.2x, when running PageRank on {UK-2007}.  Giraph and GraphX  have lower performance than GraphD and Chaos, since they are implemented based on general-purpose  Hadoop and Spark, which lack some graph specific optimizations.

We propose a new distributed graph processing system named GraphH to enable high-performance big graph analytics in small clusters. GraphH is a memory-disk hybrid approach, which does not require storing all data in memory,  but maximizes the amount of in-memory data. To achieve this goal, GraphH employs three techniques:
1) \textbf{Two-Stage Graph Partitioning.} 
GraphH performs graph partitioning in two stages. In the first stage, GraphH evenly divides the graph into a set of {tiles}, each of which uses a compact data structure to organize assigned edges. 
In the second stage, GraphH uniformly assigns {tiles} to computation servers for running vertex-centric programs.
2) \textbf{GAB (Gather-Apply-Broadcast) Computation Model.} 
GraphH uses GAB to represent distributed out-of-core graph computation. 
During the computation, each vertex maintains a replica on all servers\footnote{
A typical commodity server  can easily fit all vertices in memory. 
Take PageRank as an example, EU-2015 needs 21GB memory to store all rank values, out-degrees and intermediate messages in a single node.  Meanwhile, each server has 128GB memory  in our testbed , and has 256GB memory in \cite{shun2013ligra},\cite{wu2015g}. A single EC2 M4 instance  can have up to 256GB memory.}, 
and each computation worker  loads a tile into memory for processing at a time. GraphH uses  three functions to update a vertex:
 \emph{Gather} data along in-edges from local memory to compute an accumulator, 
 \emph{Apply} the accumulator to the target vertex,
 and \emph{Broadcast} new vertex values to other servers.  
3) \textbf{Edge Cache Mechanism.} We build an edge cache system to leverage idle memory to reduce disk I/O overhead. 



We implement GraphH using C++, MPI, OpenMP, and ZMQ. MPI is used to parallelize vertex-centric computation across multiple servers. OpenMP parallelizes the computation across multiple CPU cores of a single server. 
To  improve the communication performance, we use ZMQ to implement a broadcast interface instead of using  MPI\_Bcast.
Extensive valuations showed that GraphH performs better than existing distributed in-memory and out-of-core systems:
1) When processing generic graphs like Twitter-2010, GraphH  outperforms Giraph, GraphX, PowerGraph, PowerLyra and Pregel+ by up to 7.8x.
2) When processing big graphs like {EU-2015}, GraphH outperforms  GraphD and Chaos by at least 100x.
3) GraphH's memory management strategy is efficient, it can process big graphs like {EU-2015} even on a single commodity server without disk I/O accesses.
 GraphH is available at {https://github.com/cap-ntu/GraphH}.

The rest of the paper is structured as follows. In section II, we present the background of distributed graph processing.  Section III describes GraphH system design and implementation details. Section IV shows three performance optimization techniques. The evaluation results are detailed in Section V. Section VI concludes the paper.

\section{Background: Distributed Graph Processing}

Pregel and GAS are two widely-used  vertex-centric  models to represent distributed graph processing. 
In this section, we review 2  in-memory systems (Pregel+ and PowerGraph) and 2 out-of-core systems (GraphD and Chaos).  
Pregel+ and GraphD are designed based on the Pregel model. PowerGraph and Chaos adopt the GAS model.

\subsection{Notations}

The input graph $G=(V,E)$ has $|V|$ vertices and $|E|$ edges.  All graphs are directed graphs in this paper, and it is easy to map an undirected graph to a directed graph.
Each vertex $v \in V$ has a unique ID $id(v)$, an {incoming} adjacency list $\Gamma_{in}(v)$ and an {outgoing} adjacency list $\Gamma_{out}(v)$.  
Vertex $v$ maintains a value $val(a)$, which may be updated during the  computation, and a boolean field $active(v)$ indicating whether $v$ is active or halted.
The in-degree and out-degree of $v$ are denoted by $d_{in}(v)$ and $d_{out}(v)$, where $d_{in}(v) = |\Gamma_{in}(v)|$ and $d_{out}(v) = |\Gamma_{out}(v)|$.
If a vertex $u \in \Gamma_{in}(v)$, there is an edge $(u,v) \in E$. In this case, $(u,v)$ is an in-edge of $v$, and $u$ is an incoming neighbor of $v$. 
Similarly, if $u \in \Gamma_{out}(v)$, $(v,u)$ is an out-edge of $v$, and $u$ is an outgoing neighbor of $v$. 
Let  $val(u,v)$ denote the edge value of  $(v,u)$.
If $G$ is a unweighted graph, $val(u,v)=1, \forall (u,v) \in E$. 
The vertex state of $v$, denoted by $state(v)$, contains different components in various systems.  To be consistent, we define $state(v)$ as follows:
\begin{equation}
state(v) = (id(v), val(v), d_{out}(v), d_{in}(v), active(v)).
\end{equation}

\subsection{Graph Partitioning}

Before the vertex-centric computation, a distributed graph processing system first divides the input graph into partitions, and assigns them to $N$ servers. Figure \ref{Fig: Models} shows the graph partitioning strategies used in Pregel+, PowerGraph, GraphD and Chaos. 

\subsubsection{\textbf{Hash-based Edge-Cut Graph Partitioning}} 

Pregel+ and GraphD use a hash function to assign vertex $v$ and its outgoing adjacency list $\Gamma_{out}(v)$ to a server, and provide interaction between vertices along out-edges. Each server approximately maintains $|V|/N$ vertex states in memory. 
To enable in-memory computation, Pregel+ maintains $\Gamma_{out}(v)$ in memory, so it requires additional memory to store $|E|$ edges during the computation.
As a comparison, GraphD stores $\Gamma_{out}(v)$ on disks to reduce memory footprint during the computation. 
This hash-based edge-cut graph partitioning strategy can evenly distribute vertices among servers, 
but cannot balance workloads when processing skewed graphs, since high-degree vertices need more execution time \cite{gonzalez2012powergraph}, \cite{chen2015powerlyra}.

\subsubsection{\textbf{Intelligent Vertex-Cut Graph Partitioning}} 

PowerGraph could evenly assign $|E|$ edges to $N$ servers to improve workload balance. Specifically, if a server has edge $(u, v)$, it also maintains $state(u)$ and $state(v)$ in memory. A single vertex may have multiple replicas on different servers. For example,  both server-A and server-B in Figure \ref{Fig: Models} (b) maintain a replica of vertex-1.
Since PowerGraph requires each vertex $v$ to be aware of $\Gamma_{in}(v)$  and $\Gamma_{out}(v)$,
it needs double spaces to store an edge, which is  indexed by its source and target vertex separately.
Therefore,  PowerGraph maintains $M|V|$ vertex states and $2|E|$ edges in memory, where $M$ is the average vertex replication factor. To reduce the storage and communication overhead, 
many intelligent vertex-cut methods have been proposed to reduce the value of $M$ in PowerGraph and other GAS-based systems \cite{gonzalez2012powergraph}, \cite{chen2015powerlyra}.

\setlength{\minipagewidth}{0.45\textwidth}
\setlength{\figurewidthFour}{\minipagewidth}
\begin{figure} 
    \centering
    \begin{minipage}[t]{\minipagewidth}
    \begin{center}
    \includegraphics[width=\figurewidthFour]{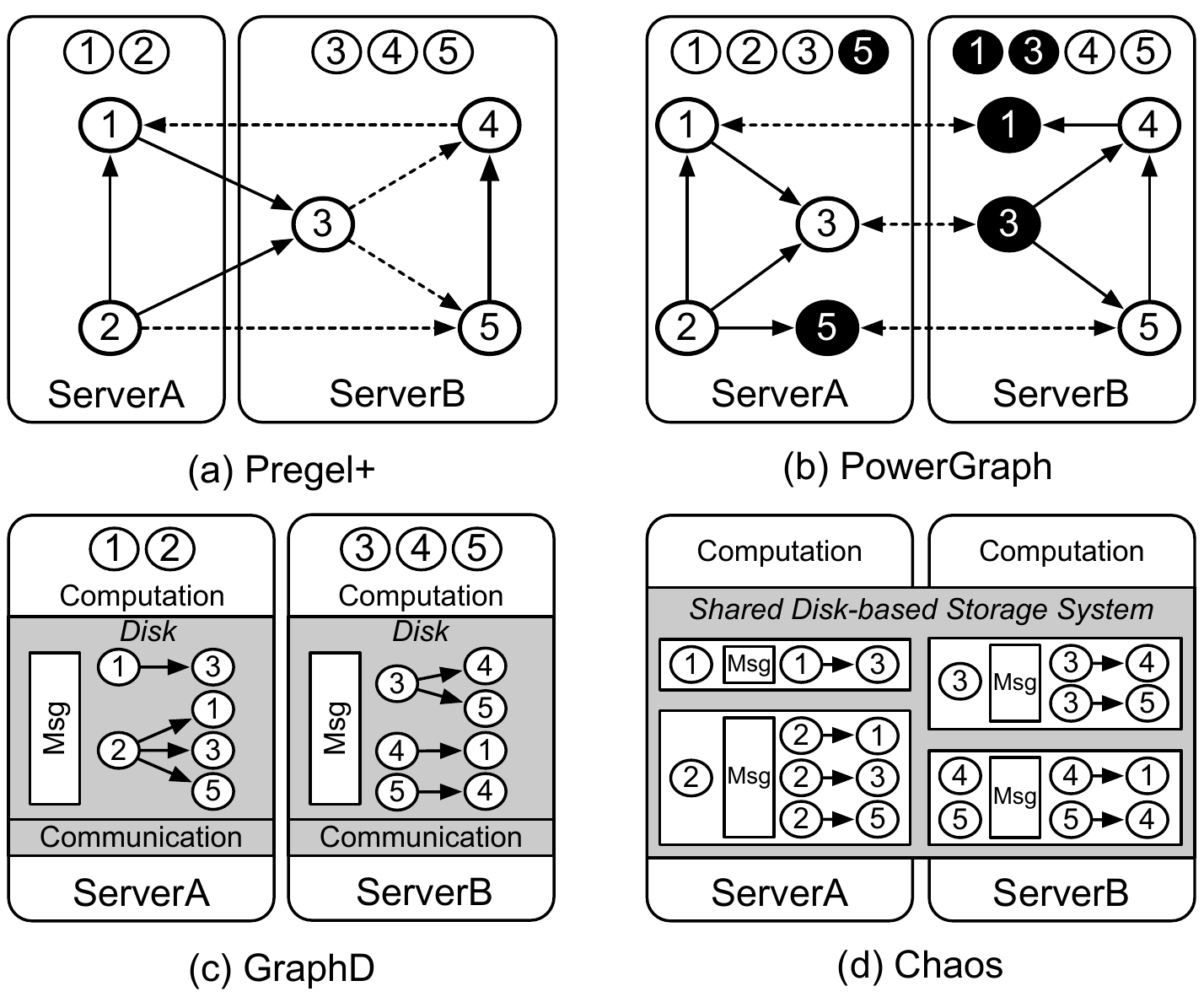}
    \end{center}
    \end{minipage}
    \centering
    \caption{Graph partitioning strategies used in Pregel+, PowerGraph, GraphD and Chaos. In (a)(c), vertices are distributed to different servers based on a hash function. In (b), the black cycle denotes a vertex replica.}
\label{Fig: Models}
\end{figure}

\subsubsection{\textbf{Streaming Partitioning}} 

Chaos divides the input graph into $P$ streaming partitions, and store them on disks.  Each partition consists of a set of vertices along with their out-edges and received messages.
All edges with the same source vertex appear in a single partition, and  they are not required to be sorted or grouped.
During the computation, each server processes a streaming partition at a time: it loads the vertices into memory and streams other data from disks. Therefore, each server only needs to maintain $|V|/P$  vertex states in memory. Chaos does not manages a streaming partition on a single server. Instead, it spreads all data of a single partition over all servers in the cluster uniformly and randomly.

\subsection{Computation Engine \& Programming Abstraction}

Pregel+ and GraphD follow the Pregel computation model to process input graphs. PowerGraph uses the GAS model to optimize the  processing of skewed graphs. Chaos designs an edge-centric GAS model to represent distributed out-of-core graph processing.

\subsubsection{\textbf{Pregel}} 

A Pregel program processes the input graph in supersteps.
In each superstep, all active vertices execute a user-defined function \emph{compute(msgs)} to update their values, then send messages along its out-edges, and vote to halt.
A halted vertex will be reactivated if it receives messages from other vertices. 
The program terminates when there are no active vetices.
To reduce the communication overhead, Pregel+ and GraphD can combine messages with the same target vertex into a single one.
Take PageRank as an example, after message combining, Pregel+ stores $\eta|E|$ and $|V|$  messages in memory at sender and receiver side respectively, where $0<\eta\leq1$ is the combining ratio\footnote{Pregel+ and GraphD only combine messages managed by the same worker. According to \cite{malewicz2010pregel}, \cite{yan2015effective}, $\eta \approx (1-exp(\frac{-d_{avg}}{TN}))\frac{TN}{d_{avg}}$, where $T$ is the number of workers in a server, $d_{avg}$ is the input graph's average degree. For example, when running PageRank on EU-2015 ($d_{avg}=85.7$) in a 9-node cluster with 216 workers, $\eta$ is excepted to be $0.82$.}. GraphD stores $|E|$ messages on disk at sender side, sends $\eta|E|$ messages over network after message combining, and digests all incoming messages in a small memory buffer. 

\begin{algorithm}
\caption{Pregel  Abstraction}\label{Alg: Pregel}
{super\_step} $\gets$ 0  \\
\While  {active\_vertices $\neq \varnothing$} {
    \For {v $\in$ active\_vertices} {
        {v.value $\gets$ v.compute(msgs)} \\ 
        v.send\_message($\Gamma_{out}$(v)), {v.halt()} 
    }
    {super\_step $\gets$ super\_step + 1} 
}
\end{algorithm}

\begin{algorithm} 
\small
\caption{GAS  Abstraction}\label{Alg: GAS}
{super\_step $\gets$ 0} \\
\While  {active\_vertices $\neq \varnothing$} {
    \For {v $\in$ active\_vertices} {
        \tcc{pull data from replicas via network}
        {accumulator $\gets$ sum(v.gather($\Gamma_{in}$(v))} \\
        \tcc{sync update to replicas via network}
        {v.value  $\gets$ v.apply(accumulator, val(v))} \\
        {v.scatter($\Gamma_{out}$(v))}
    }
    {super\_step $\gets$ super\_step + 1}
}
\end{algorithm}

\subsubsection{\textbf{GAS}}

GAS represents vertex-centric computation with three phases: \emph{gather}, \emph{apply} and \emph{scatter}. 
During the \emph{gather} phase, each active vertex collects information along its in-edges to compute an accumulator.
Since a single vertex may have multiple replicas, the \emph{gather} function runs locally on each replica, and generates a partial result. Each mirror sends its partial result to the master, which would compute the final accumulator.
In the \emph{apply} phase, the master vertex updates its values, and sends its new value to all mirrors. 
In the \emph{scatter} phase, each vertex activates its outgoing neighbors. 
Take PageRank as an example, PowerGraph sends $2M|V|$ messages via network in each superstep, where $M$ is the average vertex replication factor, and keeps $M|V|$  messages in memory during the \emph{gather} and \emph{apply} phases.

\begin{algorithm}
\caption{Edge-Centric GAS  Abstraction}\label{Alg: ECGAS}
super\_step $\gets$ 0 \\
\While  {not done} { 
\For    {p $\in$ streaming\_partitions} {
    load\_to\_memory(p.vertices) \\
    \For {e $\in$ p.out\_edges} {
      scatter(e.src.value, e.target)
    }
 }
 \For    {p $\in$ streaming\_partitions} {
    load\_to\_memory(p.vertices) \\
    \For {m $\in$ p.messages} { 
      m.target.accum $\gets$ gather(m.target, m.value) \\
    }
     \For {v $\in$ p.vertices} {
      v.value $\gets$ apply(v.accum, v.value)
    }
}
super\_step $\gets$ super\_step + 1 
}
\end{algorithm}

\subsubsection{\textbf{Edge-Centric GAS}}  
Chaos leverages an edge-centric GAS model to represent out-of-core graph processing with three phases: \emph{scatter}, \emph{gather} and \emph{apply}. 
During the \emph{scatter} phase, the \emph{Scatter} function scans all edges from streaming partitions. When dealing with a streaming partition, Chaos first loads its vertex states into memory, and process its edges sequentially. 
For each edge, the \emph{Scatter} function computes and sends a message to the target vertex. Each message is written into the associate streaming partition on disks. In the \emph{gather} phase, the \emph{Gather} function scans all messages in sequence, which are generated in the \emph{scatter} phase. For each message, the \emph{Gather} function updates the accumulator of its target vertex. Finally, in the \emph{apply} phase, the \emph{Apply} function scans all vertices and update their values sequentially. For each vertex, the \emph{Apply} function uses the corresponding accumulator to its vertex value. Take PageRank as an example,  in a single superstep, Chaos reads $2|V|$ vertex states, $|E|$ edges and $|E|$ messages from disks, and writes $|E|$ messages and $|V|$ vertex states into disks. All I/O operations need network communication, since Chaos distributes a single streaming partition across the cluster uniformly and randomly.

\renewcommand\arraystretch{1.3}
\begin{table}
\centering
\caption{Distributed Graph Processing System Comparison.}
\resizebox{0.495\textwidth}{!}{
\begin{threeparttable}
\begin{tabular}{|@{}c@{}|@{}c@{}|@{}c@{}|@{}c@{}|@{}c@{}|@{}c@{}|@{}c@{}|}
\hline 
\multicolumn{2}{|c|}{} & \multicolumn{2}{c|}{\textbf{In-Memory}} & \multicolumn{2}{c|}{\textbf{Out-of-Core}} & \textbf{Hybrid}  \\ \hline 
\multicolumn{2}{|c|}{\textbf{System}} & {Pregel+}  &  {PowerG.}  & {GaphD}  & {Chaos}  &  {GraphH}  \\ \hline 
\multirow{3}{*}{\textbf{RAM}} & {Vertex} & $\;\;\, O(|V|) \;\;\,$ & $ \;\; O(M|V|)\tnote{\#} \;\;$ & $ \;\; O(|V|) \;\;$ & $\;\, O(N|V|/P)\tnote{$\dagger$}  \;\,$ & $ O(N|V|) $ \\ 
 & {Edge} & $O(|E|)$ & $O(2|E|)$ &  $O(1)$ &  $O(1)$  & $\,O(N|E|/P)\,$\\ 
 & {Msg} & $O(\eta|E|$+$|V|)\tnote{*}\;\,$ &$ O(M|V|)$ & $O(1)$ & $O(1)$ & $O(N|V|)$ \\ \cline{1-2} 
\multicolumn{2}{|@{}c@{}|}{\textbf{Network}} & $O(\eta|E|)$ & $O(2M|V|)$ & $O(\eta|E|)$ & $O(3|E|$+$3|V|)$  & $O(N|V|)$ \\ 
\multicolumn{2}{|@{}c@{}|}{\textbf{Disk Read}} & $-$ & $-$ & $O(2|E|)$ & $O(2|E|$+$2|V|)$ & $\;O(\beta|E|)\tnote{$\ddagger$}\;$ \\ 
\multicolumn{2}{|@{}c@{}|}{\textbf{Disk Write}} & $-$ & $-$ & $O(|E|)$ & $O(|E|$+$|V|)$ & $-$ \\ \hline
\end{tabular}
\begin{tablenotes}
\footnotesize
\item[*] $\eta$ is the message combination ratio in Pregel+ and GraphD, $0 < \eta \leq 1$.
\item[\#] $M$ denotes the number of vertex replicas in PowerGraph.
\item[$\dagger$] $N$ is the server number, $P$ is the graph partition number, and $N \leq P$.
\item[$\ddagger$] $\beta$ represents the cache miss ratio in GraphH's edge cache system, $0 \leq \gamma \leq 1$.
\end{tablenotes}
\end{threeparttable}
\label{Tab: System Compare}
}
\end{table}

\subsection{Comparison}

Take PageRank as an example, Table \ref{Tab: System Compare} shows memory usage, network traffic and disk I/Os of Pregel+, PowerGraph, GraphD and Chaos. 
We should note that $|E|$  is typically much large than $|V|$ \cite{yan2016efficient}. For example, a user  could easily have tens of friends in a social network. For this reason, to reduce memory footprint, GraphD and Chaos leave all or a part of vertex states in memory, and stream edges from disks. Moreover, GraphD and Chaos need to store network-transmitted messages on disks at sender or receiver side.

Our proposed system,  GraphH,  does not require to store all data in memory, but tries to maximize the amount of in-memory data.  First, we design a GAB  model, which allows a server to process a small partition of edges in memory at a time. Therefore, the memory space required by edges is $O(N|E|/P)$, where $N$ is the server number and $P$ denotes the graph partition number.  While the GAB model stores  $|V|$ vertex states and $|V|$ messages in memory on each server, we note that current commodity servers easily fit these data in memory.  We then design an edge cache system to reduce the amount of  disk I/O operations for edges to $O(\beta|E|)$, where $\beta$ is the cache miss ratio.

\section{GraphH: System Design}

In this section, we introduce GraphH system architecture. Then, we present the graph partitioning strategy.
Next, we introduce the GAB computation model and a set of system optimizations. 
Finally, we  analyze the cost of GraphH.

\subsection{System Architecture} 

Figure \ref{Fig: Arch} shows GraphH  architecture. GraphH consists of  a distributed file system (DFS), a Spark-based graph pre-processing engine (SPE), and an MPI-based graph processing engine (MPE). 
{Tile} is the basic graph processing unit.

\subsubsection{\textbf{DFS}}
DFS centrally manages  all raw input graphs (e.g., edge lists), partitioned graphs (i.e., tiles), and  processing results. GraphH can work in conjunction with several widely used DFSs, including HDFS  and Lustre. 

\subsubsection{\textbf{SPE}}
SPE leverages Spark  to split a raw graph into disjoin sets of edges, called {tiles}, and writes them to DFS. 

\subsubsection{\textbf{MPE}}
MPE takes {tiles} as input, and runs user-defined functions on them in supersteps (or iterations). Specifically, MPE uses MPI to parallelize the computation across multiple servers in a cluster, and leverages OpenMP to parallelize the computation across multiple workers in a  server.  Each vertex has a replica on all servers. Each worker loads a tile into memory for processing at a time. In this way, MPE can handle big graphs in a small cluster with limited memory. To reduce disk I/O overhead, we build an edge cache system on each server. 
To reduce  communication overhead, we design a hybrid communication channel.

GraphH performs graph partitioning in two stages. In the first stage, SPE evenly splits the input graph's edges into $P$ {tiles}.
In the second stage, MPE uniformly assigns $P$ {tiles} to $N$ servers before running specific vertex-centric programs. SPE can be called one time for each input graph, since the pre-processing results (i.e., tiles) are persisted into DFS, and can be reused by MPE to run  many vertex-centric programs. 
 
\setlength{\minipagewidth}{0.48\textwidth}
\setlength{\figurewidthFour}{\minipagewidth}
\begin{figure} 
    \centering
    \begin{minipage}[t]{\minipagewidth}
    \begin{center}
    \includegraphics[width=\figurewidthFour]{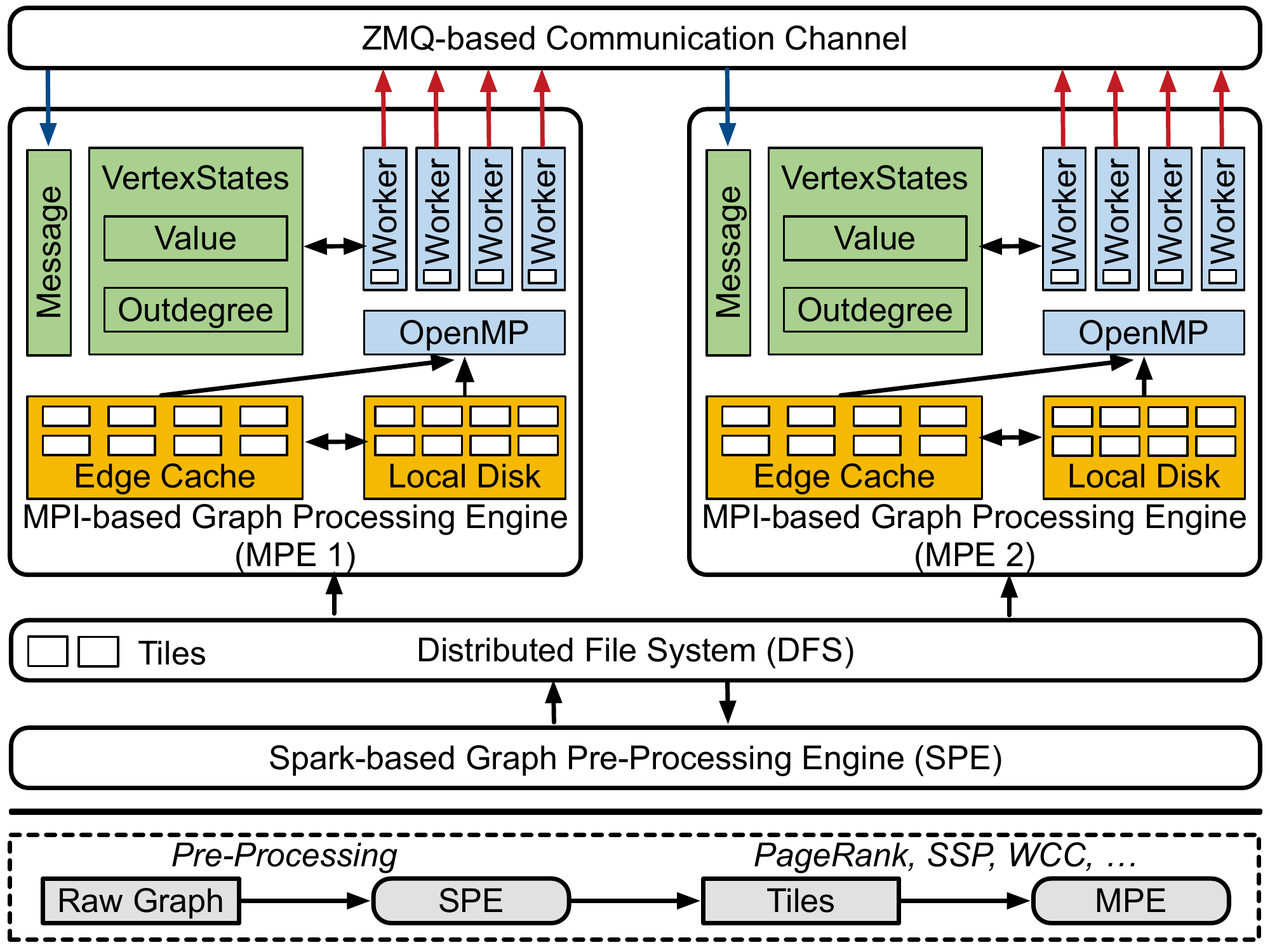}
    \end{center}
    \end{minipage}
    \centering
    \caption{GraphH system architecture.}
\label{Fig: Arch}
\end{figure}

\subsection{ Spark-based Graph Pre-Processing} 

The data pre-processing stage presents following three challenges: 1) how to evenly assign the input graph's $|E|$ edges to $P$ tiles; 2) how to organize assigned edges in a tile; and 3) how to pre-process big graphs with hundreds of millions of edges. To tackle with these challenges, we design a Spark-based graph pre-processing engine in GraphH.

\subsubsection{\textbf{Data Pre-Processing Overview}}
SPE uses three steps to split the input graph $G$ into $P$ {tiles}. 
As shown in Figure \ref{Fig: Partition_1}, in the first step, we use a $|V| \times |V|$ sparse matrix to represent $G$, where the entry in column $i$  and row $j$ is the value of edge $(i,j)$ in $G$. If $G$ is an unweighted graph, then $val(i,j)=1$. In the second step, SPE splits the sparse matrix into $P$ {tiles} in a 1D fashion, each of which roughly holds  $S = |E|/P$ nonzero entries. In Figure \ref{Fig: Partition_1}, $S = 2, P = 4$. In the third step, we organize the edges of each $tile$ in the Compressed Sparse Row (CSR) format, and persist them into DFS in binary mode. After these three steps, each {tile} has following properties: 1) Each {tile} approximately contains $|E|/P$ edges; 2) Edges appear in the same {tile} as their target vertex; 3) The target vertices in the same {tile} have  consecutive ids.
Moreover,  SPE also computes each vertex's in-degree and out-degree, and store them as two arrays in DFS.

\subsubsection{\textbf{Data Structure}}
Each {tile} organizes its assigned edges in an enhanced CSR format. Given a sparse matrix, its basic CSR format consists of three arrays: \texttt{val},  \texttt{col} and \texttt{row}. More specifically, \texttt{val} and \texttt{col}  store all nonzero entries and their column indices in row-major order, respectively. The array \texttt{row} records each vertex's edge distribution: \texttt{row[i]} and \texttt{row[i+1]}  indicate the beginning and ending offsets of vertex \texttt{i}'s column indices and edge values.
If the input graph is unweighted, all its edge values are 1, and its {tiles} would not manage the array \texttt{val} to save storage spaces.



\setlength{\minipagewidth}{0.48\textwidth}
\setlength{\figurewidthFour}{\minipagewidth}
\begin{figure}
    \centering
    \begin{minipage}[t]{\minipagewidth}
    \begin{center}
    \includegraphics[width=\figurewidthFour]{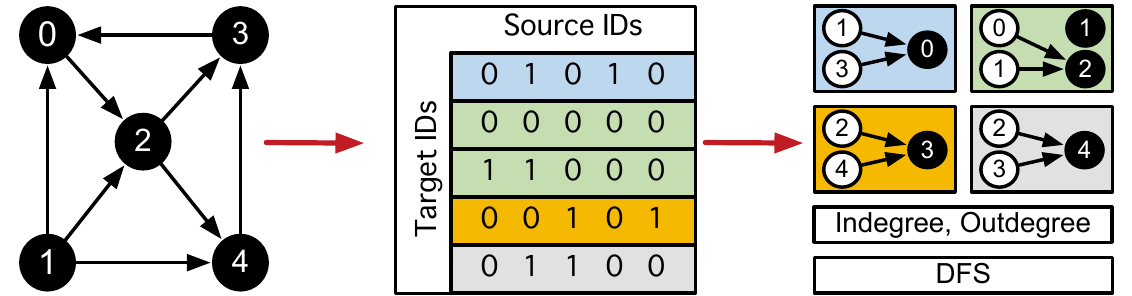}
    \end{center}
    \end{minipage}
    \centering
    \caption{SPE splits the input graph into {tiles}, each of which contains a similar number of edges in CSR format. Edges  appear in the same {tile} as their target vertex. SPE also stores $d_{in}(v)$ and  $d_{out}(v)$ for each vertex $v$.}
\label{Fig: Partition_1}
\end{figure}

\begin{algorithm} 
\small
\caption{Spark-based Data Pre-Processing}\label{Alg: Spark_Split}
outdegree = edges.map(e $\Rightarrow$ (e.src, 1)).reduce(SUM) \\
indegree  = edges.map(e $\Rightarrow$ (e.target, 1)).reduce(SUM) \\
\BlankLine
tile\_id = 0, size = 0, splitter = empty   \\
\While {vertex\_id $<$ vertex\_number} {
   size +=  indegree[vertex\_id], splitter[tile\_id+1] = vertex\_id \\
  \If (\tcp*[f]{S is avg tile size}){ size $\geq$ S}  { 
    tile\_id += 1, size = 0 \\
  }
  vertex\_id = vertex\_id + 1
} \BlankLine
kv = edges.map(e $\Rightarrow$ (get\_tile\_id(e.target, splitter), e)) \\
tiles = kv.group\_by\_key().to\_CSR() \\
save\_to\_dfs(outdegree, indegree, tiles) \\
\end{algorithm}

\subsubsection{\textbf{Tile Size}}
GraphH allows users to manually configure the average tile size $S$ (i.e.,  number of edges in a {tile}), where $S = |E|/P$.
Since {tile} is the basic processing unit on MPE, $S$ has high impact on performance.
If $S$ is too large, GraphH may have the memory overflow problem. For example, if $S=256\text{M}$, each {tile} consumes more than $1$GB memory. Given a server with $24$ workers, GraphH must reserve at least $24$GB memory for {tiles}. 
Since the  power-law distribution of vertex degrees can be observed in most real-world graphs \cite{gonzalez2012powergraph}, if $S$ is too small, the size of  a {tile} would be bounded by high-degree vertices. As a result, GraphH cannot evenly split the input graph into {tiles}, leading to storage and computation imbalance.
In this paper, we configure $S$ to be a value between $15$M and $25$M to balance the storage and computation requirements.

\subsubsection{\textbf{Splitting Big Graphs on Spark}}
SPE replies on Spark to pre-process big graphs using three map-reduce jobs, as shown in Algorithm \ref{Alg: Spark_Split}. The first two map-reduce jobs (line 1-2) return each vertex's in-degree and out-degree. 
Then, we traverse the in-degree array, assign all encountered vertex's in-edges to a {tile} until it has more than $S = |E|/P$ edges, and store the split information in a \emph{splitter} array. Given a vertex $v$, all its in-edges are assigned to {tile[t\_id]}, if {\emph{splitter}[t\_id] $\leq$ $v$ $<$ \emph{splitter}[t\_id+1]}. We run the third map-reduce job (line 9-10) to group edges by their tile ids, and organize them in CSR.

\renewcommand\arraystretch{1.1}
\begin{table} [H]
\centering
\resizebox{0.48\textwidth}{!}{
\begin{threeparttable}{}
\caption{Input data size (GB) for different graph processing systems.}
\begin{tabular}{|c|c|c|c|c|c|}
\hline
\textbf{Graphs} & \textbf{\begin{tabular}[c]{@{}c@{}}Edge List\\ (CSV) \end{tabular}} & \textbf{\begin{tabular}[c]{@{}c@{}}Pregel+\\ GraphD\end{tabular}} & \textbf{Giraph} & \textbf{Chaos} & \textbf{GraphH\tnote{\#}} \\ \hline 
\textbf{Twitter-2010} & 24 & 12 & 18 & 11 & 7 \\ 
\textbf{UK-2007} & 94 & 48 & 69 & 38 & 25 \\ 
\textbf{UK-2014} & 874 & 445 & 624 & 351 & 204 \\
\textbf{EU-2015} & 1700 & 862 & 1220 & 684 & 378 \\ \hline
\end{tabular}
\label{Tab: TileSize}
\begin{tablenotes}
\footnotesize
\item[\#] Including all tiles, vertex in-degree array and vertex out-degree array.
\end{tablenotes}
\end{threeparttable}
}
\end{table}

\subsubsection{\textbf{SPE Output}}
SPE converts the input graph into  tiles along with an in-degree array and an out-degree array, and use them as the input of specific applications on MPE.
We show the input data size of MPE and other graph processing systems in Table \ref{Tab: TileSize}. GraphX, PowerGraph and PowerLyra  use edge list as input.
Pregel+, GraphD, Giraph and Chaos require users to manually convert the input graph to a given format.
We observe that SPE also reduces the storage space significantly, other than evenly splitting the input graph.
For example, GraphH needs 378GB disk to store EU-2015 in tiles, while its raw edge list needs 1.7TB.  The corresponding values of Pregel+, Giraph and Chaos are 862GB, 1.22TB,and 684GB, respectively.
Vertex-centric programs can benefit a lot from this compact graph representation: 1) reduced disk I/O overhead when accessing on-disk edges; 2) improved the amount of edges that can be cached in memory.

\subsection{GAB-based Vertex-Centric Computation} 

We design an MPI-based graph processing engine (MPE) to perform vertex-centric computation. 
We first describe the data layout in MPE. Next, we present the GAB computation model and the parallel computation strategy.
The complete pseudo-code description of MPE is shown in Algorithm \ref{Alg: GAB}.

\subsubsection{\textbf{Data Layout}} When running a vertex-centric program on MPE, each server maintains three types of data: tiles (partitioned edges), vertex  states  and  messages. MPE uses following strategies to manage these data.

\begin{itemize}

\item \emph{Tiles.} Each server stores all assigned tiles on local disk, and uses an edge cache system to store a portion of tiles in memory. 
MPE uniformly assigns $P$ tiles across $N$ servers as following: $i$th tile is assigned to $j$th server if $i \pmod N = j$, and make each server fetch assigned tiles from DFS to local disk.  

\item \emph{Vertices.} Each vertex has a replica on all servers. Thus, each server stores $|V|$ vertex states in memory. GraphH leverages a list of dense arrays to represent vertex states, and allow users to decide which array to include. For example, each server maintains a rank value array and an out-degree array to run PageRank. 

\item \emph{Messages.} GraphH stores all messages in memory. As shown in Algorithm \ref{Alg: GAB} (line 13), a server only transmits updated vertex values to other servers. Therefore, each server uses a $|V|$-dimensional dense array to manage all received messages.

\end{itemize}

\subsubsection{\textbf{GAB  Model}}

GraphH abstracts vertex-centric programs into the GAB (Gather, Apply, Broadcast) computation model. GAB is designed based on the GAS (Gather, Apply, Scatter) model, but differs significantly. With GAB, each vertex executes three functions to update its value in supersteps: the \emph{gather} function collects information along its in-edges and compute an accumulator; the \emph{apply} function uses the accumulator to produce an updated vertex value; the \emph{broadcast} function copies the updated vertex value to other replicas. 
The vertex-centric program terminates when there are no updated vertices. 
GAB only requires users to implement the \emph{gather} and \emph{apply} functions.

\begin{algorithm} 
\small
\caption{MPE Computation Engine}\label{Alg: GAB}
\For {tile $\in$ all\_tiles} { 
    \If {tile.id (mode N) = server\_id} {
      assigned\_tiles.append(tile) \\
    }
}
{load\_tiles\_from\_dfs\_to\_disk(assigned\_tiles)}, {initial\_vertices()} \\
\While  {updated\_vertex\_num $> 0$} {
  updated\_vertex\_num = 0; \\
  \tcc{Parallelized computation on T workers}
  \# pragma omp parallel for num\_threads(T)\\
  \For {tile $\in$ assigned\_tiles} { 
    \If {tile.bloom\_filer\_contains(updated\_vertices)} {
    {load\_to\_memory(tile)} \\
      \For {v $\in$ tile.target\_vertices} {
        {accum = \textbf{Gather}(v.in\_edges, vertex\_states)} \\
        {updated\_value = \textbf{Apply}(v.accum, v.value)} \\
        \If {updated\_value $\neq$ v.value} {
          {\textbf{Broadcast}(v.vertex\_id, updated\_value)} \\
        }
      }
      {free\_memory(tile)} \\
      }
    }
    {wait\_other\_servers()} \\
    \For {v $\in$ all\_vertices} {
      \If {v.has\_updated\_value()} {
        v.value = v.updated\_value\\
        updated\_vertex\_num += 1
      }
    }
    {super\_step = super\_step + 1} \\
}
\end{algorithm}


In Algorithm \ref{Alg: GAB-PR} and \ref{Alg: GAB-SSSP},  we use GAB to implement PageRank and single source shortest path (SSSP). In PageRank, for each vertex, the \emph{gather} function collects information along its in-edges and sums them to be an accumulator. The \emph{apply} function then produces a new rank value. 
In SSSP, the \emph{gather} function computes the shortest path through each of the in-edges, and the \emph{apply} function returns the new distance. Both PageRank and SSSP require users to implement an additional function to initiate all vertex values and specificity required vertex state components. For example, PageRank needs to load the vertex out-degree array into memory.

\subsubsection{\textbf{Parallelized Out-of-Core GAB Computation on MPE}}
MPE performs parallelized out-of-core GAB computation at the level of tiles. 
As shown in Algorithm \ref{Alg: GAB} (line 6 - 22), each server processes assigned tiles with $T$ workers in supersteps. 
During the computation,  each worker is scheduled to process a tile at a time. More specifically, a worker firstly loads a tile into memory, and then traverses it from the first target vertex. For each target vertex,  user-defined \emph{gather} and \emph{apply} functions are called sequentially to produce an update value. The \emph{gather} function would not incur any network communications, since each vertex has a replica on all servers. 
Finally, MPE calls the \emph{broadcast} function to copy the updated vertex value to all other replicas. 
In addition, MPE follows the Bulk Synchronous Parallel (BSP) synchronization model. As shown  in Algorithm \ref{Alg: GAB} (line 15 - 22), after all servers have completed tile preprocessing, each server updates its vertex replicas.

\subsubsection{\textbf{Avoid Loading Inactive Tiles}}
For many algorithms,  GraphH may just update a few vertices in a superstep. If a tile does not contain any source vertices with updated values,  GraphH would not update any target vertices,  wasting time to load and process it.
To solve this problem,  GraphH makes each tile leave a bloom filter in memory to record its source vertex information. When processing a tile, GraphH would first check whether its source vertex list contains any update vertices. If yes, GraphH would continue to load the tile into memory for processing.  Otherwise, GraphH skips this tile.


\SetKwProg{Fn}{Function}{}{}
\begin{algorithm} 
\small
\caption{PageRank implemented with GAB}\label{Alg: GAB-PR}
\Fn{PageRank\_initial\_vertex\_states()}{
    load\_to\_memmory(vertices.out\_degree) \\
    \textbf{for} v $\in$ vertices \textbf{do}  v.value = 1 / num\_vertex \\
}
\Fn{PageRank\_Gather(v.in\_edges, vertices)}{
    \For {e $\in$ v.in\_edges} {
        accum += e.source.value / vertices.out\_degree[u.id] \\
    }
    \Return accum \\
}
\Fn{PageRank\_Apply(v.accum, v.value)}{
    updated\_value = 0.15 / num\_vertex + 0.85 * v.accum \\
    \Return updated\_value \\
}
\end{algorithm}

\SetKwProg{Fn}{Function}{}{}
\begin{algorithm} 
\small
\caption{SSSP implemented with GAB}\label{Alg: GAB-SSSP}
\Fn{SSSP\_initial\_vertex\_states()}{
    \textbf{for} v $\in$ vertices \textbf{do} v.value = $\infty$
    all\_vertices[source\_vertex\_id] = 0 \\
}
\Fn{SSSP\_Gather(v.in\_edges, vertices)}{
    accum = $\infty$ \\
    \For {e $\in$ v.in\_edges} {
        accum = min (e.source.value + e.value, accum)  \\
    }
    \Return accum \\
}
\Fn{SSSP\_Apply(v.accum, v.value)}{
    \Return min (v.accum, v.value) \\
}
\end{algorithm}

\setlength{\minipagewidth}{0.49\textwidth}
\setlength{\figurewidthFour}{\minipagewidth}
\begin{figure} 
    \centering
    \begin{minipage}[t]{\minipagewidth}
    \begin{center}
    \includegraphics[width=\figurewidthFour]{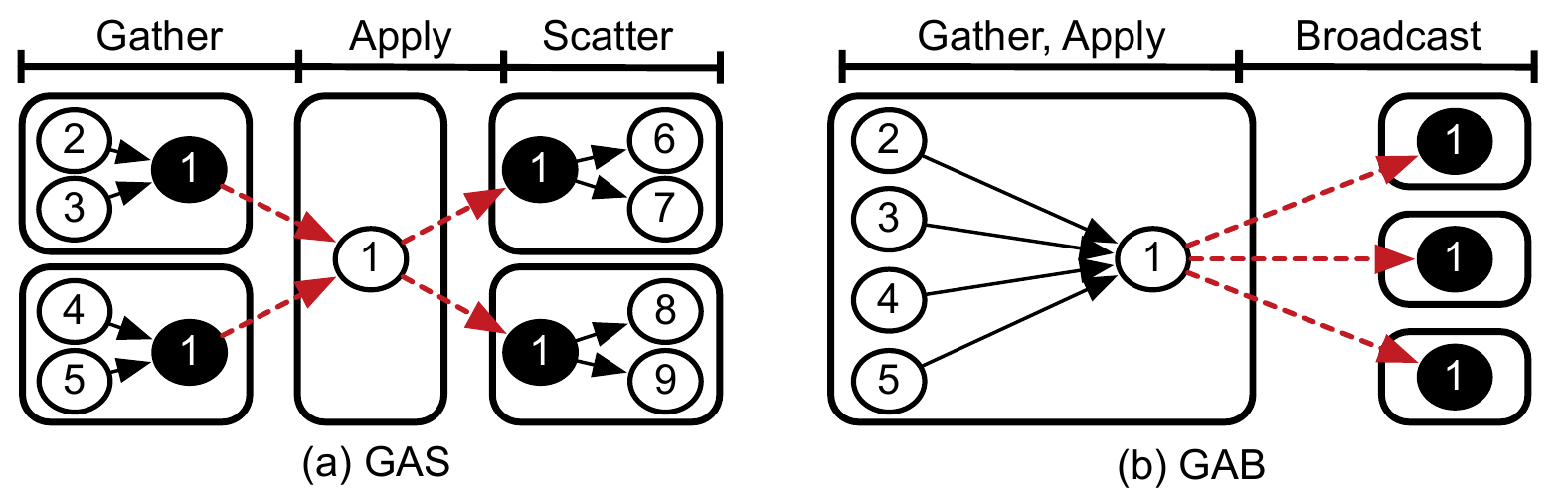}
    \end{center}
    \end{minipage}
    \centering
    \caption{GAS vs. GAB.  GAS requires thee computation operations (local sum in the \emph{gather} phase, update vertex in the \emph{apply} phase, and active neighbors in the \emph{scatter} phase) and two communication operations to update a vertex.  GAB  needs one computation operation (\emph{gather} and \emph{apply}), and one communication operation (broadcast updated vertex values).}
\label{Fig: cc}
\end{figure}

\subsubsection{\textbf{GAS vs. GAB}} 
GAB is designed based on GAS, but differs significantly, as shown in Figure \ref{Fig: cc}.  
First, GAS needs five operations to update a single vertex, and GAB  requires a computation and a communication operation. 
Second, GAS tries to reduce the vertex replication factor by placing edges intelligently  \cite{gonzalez2012powergraph}, while GAB requires each vertex to have a replica on all servers. However, GAB does not maintain all edges in memory, thus it uses much less memory than GAS. 
Third, given an updated vertex, GAS would activate its neighbors along out-edges. Only active vertices can join the computation of next superstep. 
As a  comparison, GAB does not contain this step to save the memory used to store out-edges. At each superstep, GAB make each vertex checks whether its in-coming neighbors have new values and run the \emph{gather},  \emph{apply} functions accordingly.

\section{System Optimizations}

To improve system performance, GraphH involves a set of optimizations on system storage and communication. 

\subsection{{All-in-All Vertex Replication}}

GraphH uses an \emph{All-in-All (AA)} vertex replication policy to manage vertex states across $N$ servers. Specifically, the \emph{AA} policy requires each vertex has a replica on all servers. Therefore, each server maintains all $|V|$ vertex replicas in memory, even if some of them are used. 
While the \emph{AA} policy wastes some memory to store unused vertices, it can manage each vertex state component in a dense array  without indexing overhead. For example, when running PageRank, each server maintains a value array and an out-degree array. Given a vertex $v$, its position in the array is equal to $id(v)$. Moreover, with the $AA$ policy, GraphH also manages all network-transmitted messages in a dense array, which is used to update vertex values at the end of a superstep. 
Let $M_{i, aa}$ denote the amount of memory required by GraphH for vertex-centric computation on $i$th server with the \emph{AA} policy:
\begin{equation}
{M}_{i, AA} = {Size(Vertex, Msg)} \times |V| + Size(Tile) \times T,
\end{equation}
where $Size(Vertex, Msg)$ is the size of a vertex state and a message, and $T$ denotes the number of workers in a server, each of which processes a tile in memory  at a time.

The \emph{On-Demand (OD)} replication policy can avoid storing unused vertex states, but incurs additional indexing overhead. Let $V_{i, OD}$ be the set of vertices managed in $i$th server with the \emph{OD} policy.
More specifically, $|V_{i,OD}|$ only contains the source and target vertices that appear in assigned tiles of $i$th server. Each vertex state and its updated value  are  indexed and positioned by the vertex id. With the \emph{OD} policy, the amount of required memory  in $i$th server is: 
\begin{equation} 
{M}_{i, OD} = {Size(ID, Vertex, Msg)} \times |V_{i, OD}| + Size(Tile) \times T.
\end{equation}
Assume the input graph $G$ is a random graph: the neighbors of a vertex in $G$ are randomly chosen among $V$. In this case,  $|V|$ vertices and $|E|$ edges are evenly assigned to $N$ servers. The expected number of vetices maintained by $i$th server is:
\begin{equation}
E[|V_{i, od}|]  \leq (1 - (1 - {d_{avg}}/{|V|})^{|V|/N}))|V| + {|V|}/{N}, \\
\end{equation}
where $(1 - (1 - {d_{avg}}/{|V|})^{|V|/N}))|V|$ is the number of source vertices, ${|V|}/{N}$ is the amount of target vertex number, and some target vertices may appear in the source vertex list.
For big graphs, due to $\lim_{n \to \infty}{(1-1/n)}^n = e^{-1}$, we have 
\begin{equation}
E[|V_{i, od}|]  \leq (1 - e^{-d_{avg}/N}) |V| + |V|/N.
\end{equation}

We take PageRank as an example to show that the \emph{AA} policy is more memory efficient than the \emph{OD} policy in small clusters, since it eliminates the indexing overhead at the cost of storing unused vertices and messages. Specifically, with the \emph{AA} policy,  $Size(Vertex, Msg) = 20 \text{ bytes}$, since each  vertex value and message is represent by a double-precision number (8 bytes), and each vertex out-degree is an integer (4 bytes). When using the \emph{OD} policy, $Size(ID, Vertex, Msg) = 24 \text{ bytes}$, since each vertex id is represented by an unsigned integer (4 bytes). Assume that Twitter-2010, UK-2007, UK-2014 and EU-2015 are random graphs\footnote{This assumption may not be accurate for a real graph. We use experiments to show that expected memory usage is effective on real graphs.}, Figure \ref{Fig: AllinAll} (a) shows the expected memory usage per server. We can observe that the \emph{AA} policy consumes less memory than the \emph{OD} policy for all graphs in a small cluster with less than $16$ servers. In a big cluster with more than 48 servers, the \emph{OD} policy consumes less memory than the \emph{AA} policy to run PageRank on EU-2015. Since GraphH is designed for big graph analytics in small clusters, we use the \emph{AA} policy in GraphH for memory-efficiency purpose. 

Figure \ref{Fig: AllinAll} (b) shows the memory usage of GraphH (using the \emph{AA} policy without edge cache) per server to run PageRank and SSSP in a 9-node cluster. 
We see that the \emph{AA} policy would not be a bottleneck, since current single commodity server can easily fit all vertex states and messages in the main memory. 
For example, to run PageRank on EU-2015, each server roughly consumes 33GB memory (including HDFS). The corresponding value of SSSP is 18GB. 

\setlength{\minipagewidth}{0.237\textwidth}
\setlength{\figurewidthFour}{\minipagewidth}
\begin{figure} 
    \centering
    \begin{minipage}[t]{\minipagewidth}
    \begin{center}
    \includegraphics[width=\figurewidthFour]{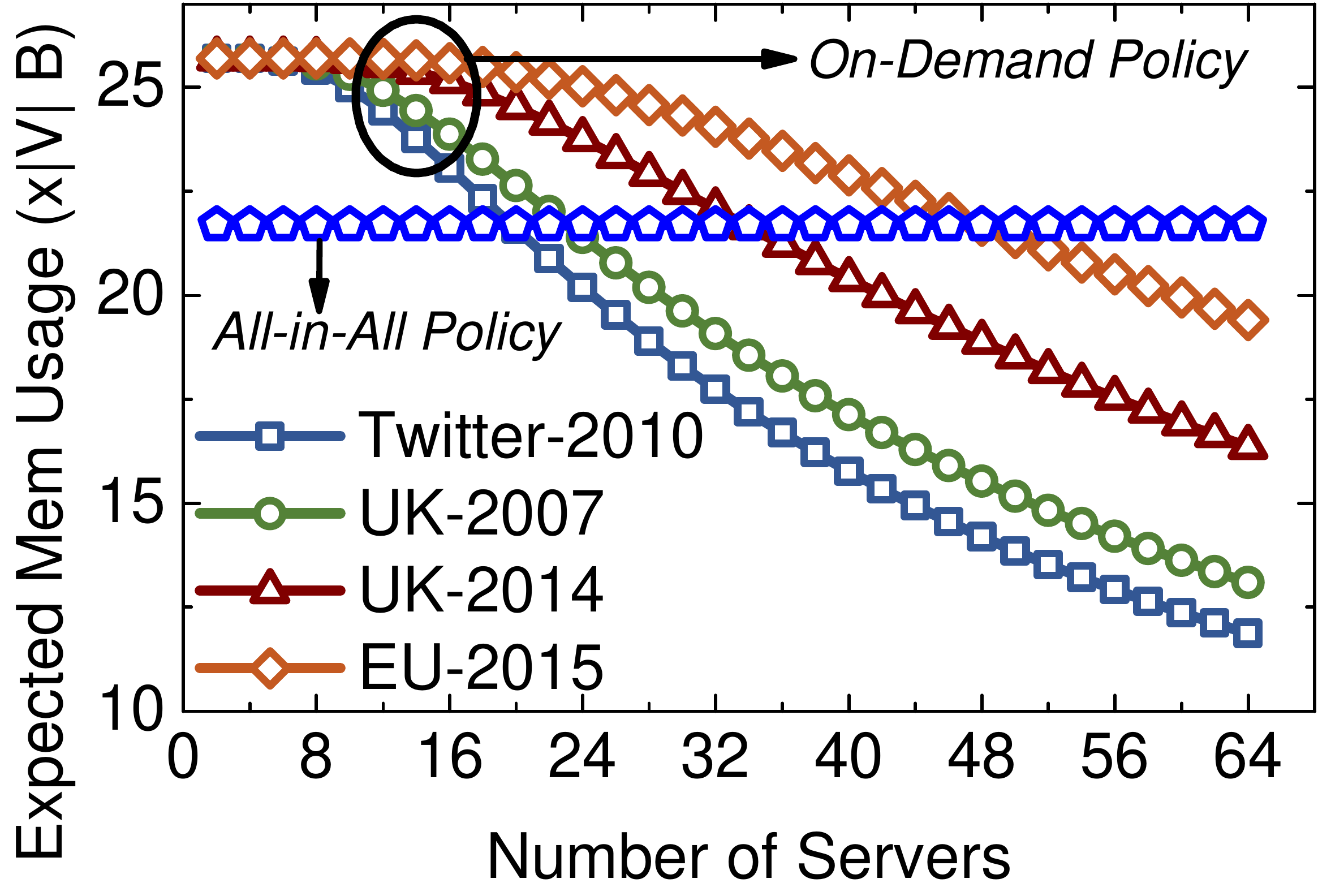}
    \subcaption{(a) Expected memory usage per server.}
    \end{center}
    \end{minipage}
    \centering
    \begin{minipage}[t]{\minipagewidth}
    \begin{center}
    \includegraphics[width=\figurewidthFour]{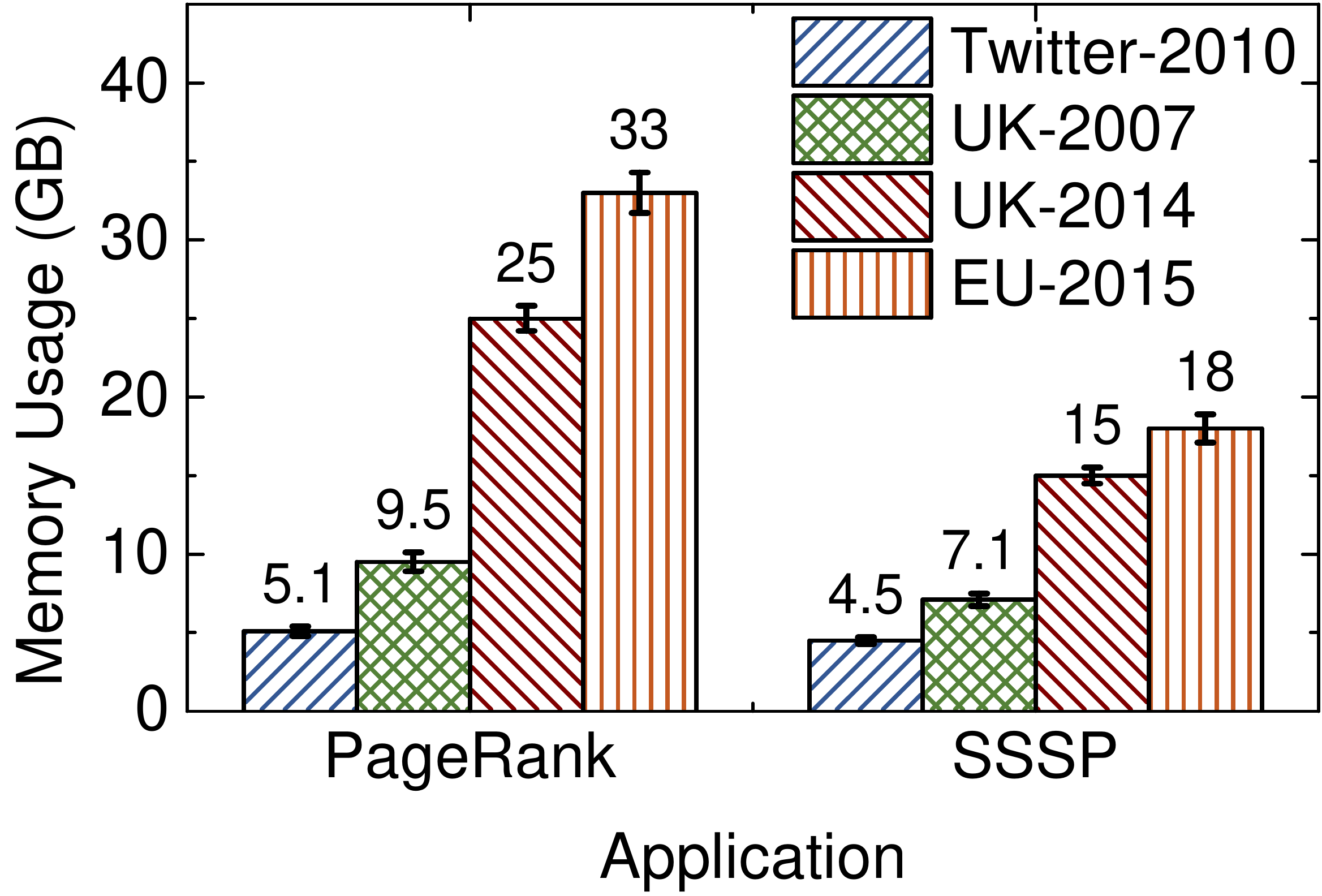}
    \subcaption{(b) Memory usage per server in testbed.}
    \end{center}
    \end{minipage}
    \centering
    \caption{GraphH memory usage analytics. (a) shows the expected amount of used memory per server, when running PageRank with the All-in-All policy and On-Demand policy. (b) shows the GraphH (using the \emph{AA} policy without edge cache) memory usage  per server in a 9-node cluster to run PageRank and SSSP. Each server in the cluster has 128GB memory.}
\label{Fig: AllinAll}
\end{figure}

\subsection{{Edge Cache Mechanism}}

We design an edge cache system  to further reduce the disk I/O overhead of GraphH.
As shown is  Figure \ref{Fig: AllinAll} (b), when running PageRank on EU-2015 in a 9-node cluster, each server only uses 33GB memory, leaving other 95GB memory idle. 
To reduce the amount of costly disk accesses, we build a tile cache system on these idle memory.
During the vertex-centric computation, when a worker needs to load a tile, it firstly searches the cache system. If hit, the worker can get the target tile without disk I/O operations. Otherwise, the worker reads the target tile from local disks, and leaves it in the cache system if the cache system is not full.

\renewcommand\arraystretch{1.1}
\begin{table}
\centering
\caption{Compression ratio and processing throughput per CPU core.}
\label{Tab: compress1}
\resizebox{0.48\textwidth}{!}{
\begin{tabular}{|c|c|c|c|c|c|c|}
\hline
& \multicolumn{3}{c|}{\textbf{Compression Ratio}} & \multicolumn{3}{c|}{\textbf{Throughput (MB/s)}} \\ \hline
& snappy  & zlib-1  & zlib-3  & snappy & zlib-1  & zlib-3           \\ \hline 
\textbf{Twitter-2010} & 1.75 & 2.78 & 3.22  & 870 & 55  & 46    \\ 
\textbf{UK-2007} & 1.89 & 3.71 & 4.54 & 947  & 58  & 53 \\ 
\textbf{UK-2014} & 1.96 & 4.34  & 5.26 & 903   & 65 & 50 \\ 
\textbf{EU-2015} & 1.96 & 4.35  & 5.88  & 890 & 62  & 56 \\ \hline
\end{tabular}
}
\subcaption{}
\resizebox{0.48\textwidth}{!}{
\begin{tabular}{|c|c|c|c|c|c|}
\hline
 & \textbf{\begin{tabular}[c]{@{}c@{}}Edge List\\ (CSV)\end{tabular}} & \textbf{\begin{tabular}[c]{@{}c@{}}Tile\\ (raw)\end{tabular}} & \textbf{\begin{tabular}[c]{@{}c@{}}Tile\\ (snappy)\end{tabular}} & \textbf{\begin{tabular}[c]{@{}c@{}}Tile\\ (zlib-1)\end{tabular}} & \textbf{\begin{tabular}[c]{@{}c@{}}Tile\\ (zlib-3)\end{tabular}} \\ \hline 
\textbf{Twitter-2010} & 24GB & 6.5GB & 3.7GB & 2.3GB & 2GB \\ 
\textbf{UK-2007} & 94GB & 23GB & 12GB & 6.2GB & 5GB \\ 
\textbf{UK-2014} & 874GB & 196GB & 100GB & 45GB & 37GB \\ 
\textbf{EU-2015} & 1700GB & 362GB & 185GB & 80GB & 62GB \\ \hline
\end{tabular}
}
\end{table}

To further reduce disk I/O overhead, GraphH can compress tiles in the edge cache system. Table \ref{Tab: compress1} shows that popular compressors, such as snappy and zlib, can efficiently reduce the data size of real-world graphs.  For example, zlib-3 ({$N$ denotes the compression level of zlib in zlib-$N$}) can compress EU-2015 tiles by a factor of $5.88$, and reduce its data size to 62GB. 
While workers need additional decompression time, GraphH's edge cache system still provides much higher performance than hard disks. Table \ref{Tab: compress1} shows that snappy can decompress tiles at a rate of up to 903MB/s per CPU core. The corresponding value of zlib-3 is 56MB/s. If a server has 22 workers, its overall tile loading rate of zlib-3  is about 1.2GB/s. In contrast, we can only achieve up to $310$MB/s sequential disk read speed with RAID5, and the available disk bandwidth is shared by all workers of a server.

GraphH's edge cache system can automatically switch to the most suitable mode,  considering disk I/O and decompression overhead. In this work, we consider 4 cache modes:
\begin{itemize}
\item Mode-1: Cache uncompressed  tiles.
\item Mode-2: Cache compressed tiles processed by snappy.
\item Mode-3: Cache compressed tiles processed by zlib-1.
\item Mode-4: Cache compressed tiles processed by  zlib-3.
\end{itemize}
When having limited memory, it is crucial to reduce the disk I/O overhead by selecting compressors or libraries with high compression rate. As shown in Figure \ref{Fig: CompressCompare}, when running PageRank on EU-2015 using three servers, compared to mode-1, mode-3 could improve the system performance by a factor of 17.6 by caching all tiles in memory. With the same cache hit ratio, the decompression overhead can reduce the system performance. For example,  Figure \ref{Fig: CompressCompare} shows that mode-4 increases the execution time by a factor of 2 with 9 servers, compared to mode-1. To minimize disk I/O overhead as well as decompression overhead, GraphH automatically selects the most suitable cache mode at the beginning a vertex-centric program.
Let $C$ denote GraphH's edge cache capability,  $S$ is the input graph's tile size, and $\gamma_i$ is the estimated compression ratio of  cache mode-$i$.  GraphH would minimize $i$ constrained by $S/\gamma_i  \leq C$. If no mode can satisfy this constraint, GraphH would use mode-3. In this work, $\gamma_0 = 1, \gamma_1 = 2, \gamma_2 = 4, \gamma_3 = 5$, according to  Table \ref{Tab: compress1}.

\setlength{\minipagewidth}{0.237\textwidth}
\setlength{\figurewidthFour}{\minipagewidth}
\begin{figure} 
    \centering
    \begin{minipage}[t]{\minipagewidth}
    \begin{center}
    \includegraphics[width=\figurewidthFour]{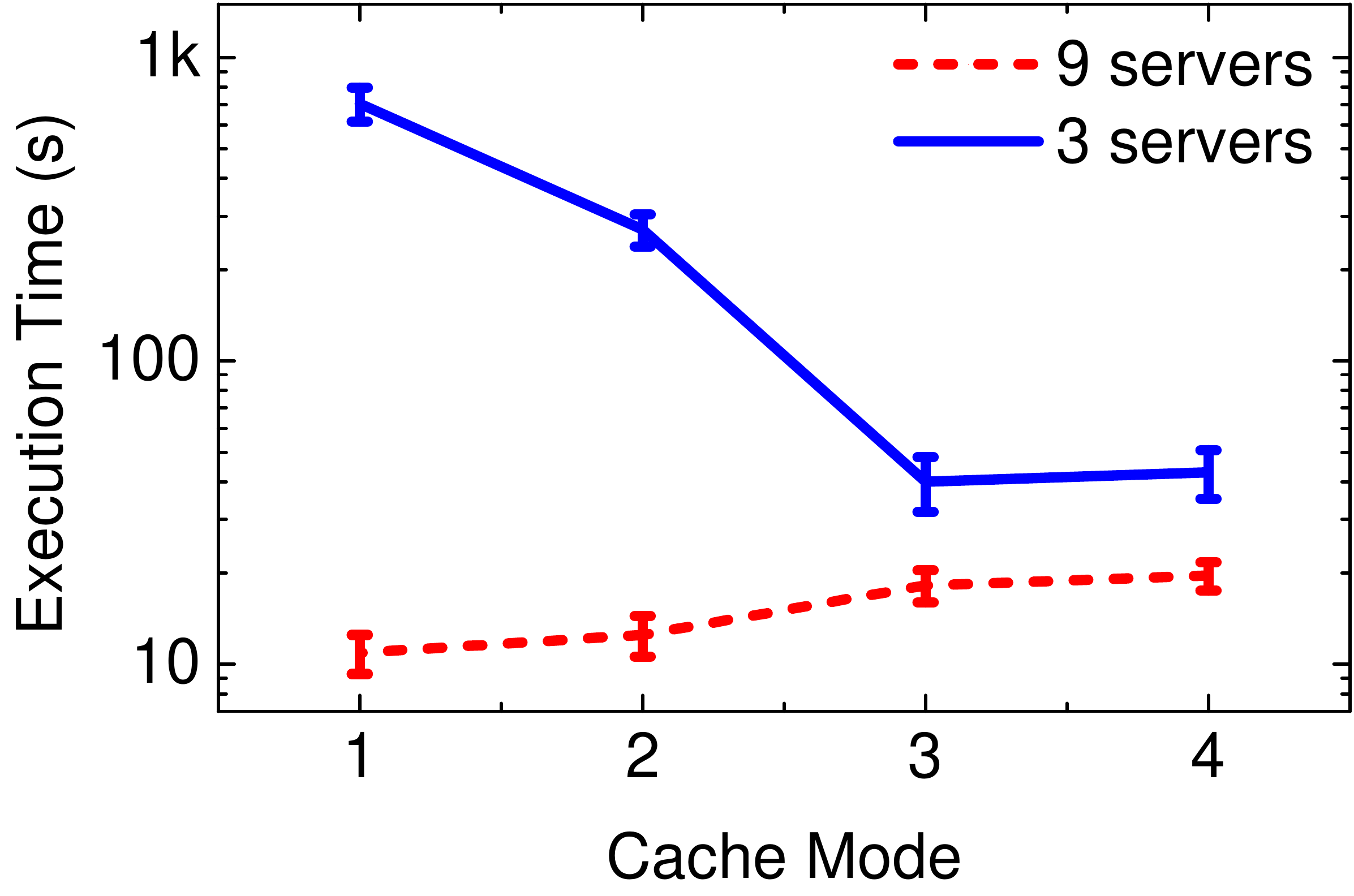}
    \subcaption{(a) Execution time per superstep.}
    \end{center}
    \end{minipage}
    \centering
    \begin{minipage}[t]{\minipagewidth}
    \begin{center}
    \includegraphics[width=\figurewidthFour]{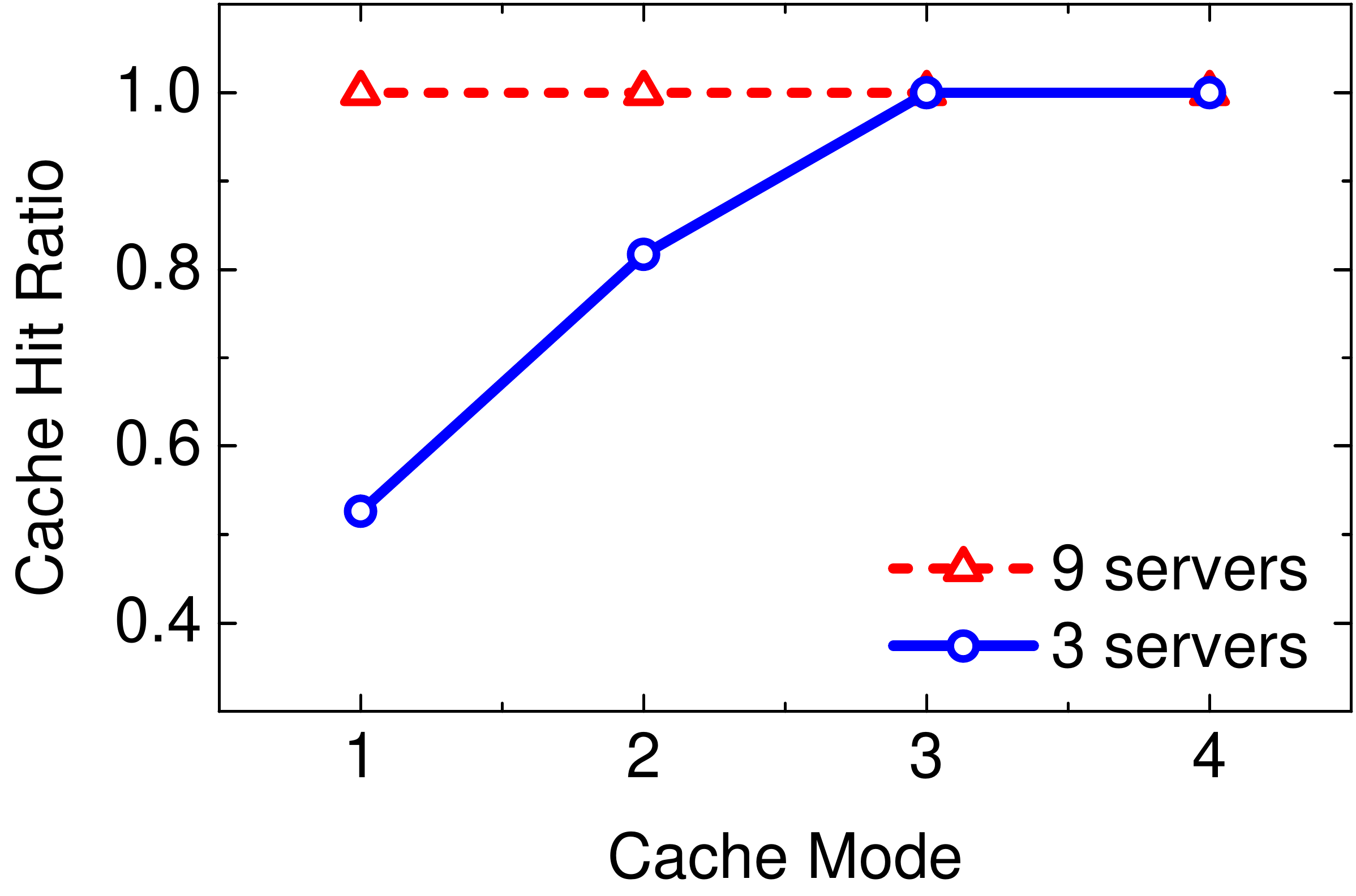}
    \subcaption{(b) Cache hit ratio.}
    \end{center}
    \end{minipage}
    \centering
    \caption{Execution time and cache hit ratio comparison with different cache modes. These experiments are done using PageRank on EU-2015.}
\label{Fig: CompressCompare}
\end{figure}

\subsection{{Hybrid Communication Mode}}

We observe that a single communication mode cannot perform well all the time for vertex-centric programs. 
More specifically,  each worker generates a set of updated vertex values when processing a tile. GraphH makes each worker  buffer updated vertex values, and broadcast them to other servers in a single message after processing  the whole tile. 
It is advantageous to use a dense array representation for updated vertex values along with a bitvector to record updated vertex id. However, this dense communication mode may waste a lot of network bandwidth when a few of vertices are updated, because it needs to send many zeros.  For example, Figure \ref{Fig: hybridnetwork} (a) shows that less than 50\% vertices are updated after the 160th superstep when running PageRank on UK-2007 in a 9-node cluster. 
The sparse array representation, which converts a dense array into a list of indices and values, can solve this problem, since it only sends updated vertex values. 
However, sparse communication mode would waste a huge amount of network resources to send indices, if the vertex updated ratio is high. As shown in Figure \ref{Fig: hybridnetwork} (b), only after the 160th superstep, the sparse mode  has lower network traffic than the dense mode. 

\setlength{\minipagewidth}{0.238\textwidth}
\setlength{\figurewidthFour}{\minipagewidth}
\begin{figure} 
    \centering
    \begin{minipage}[t]{\minipagewidth}
    \begin{center}
    \includegraphics[width=\figurewidthFour]{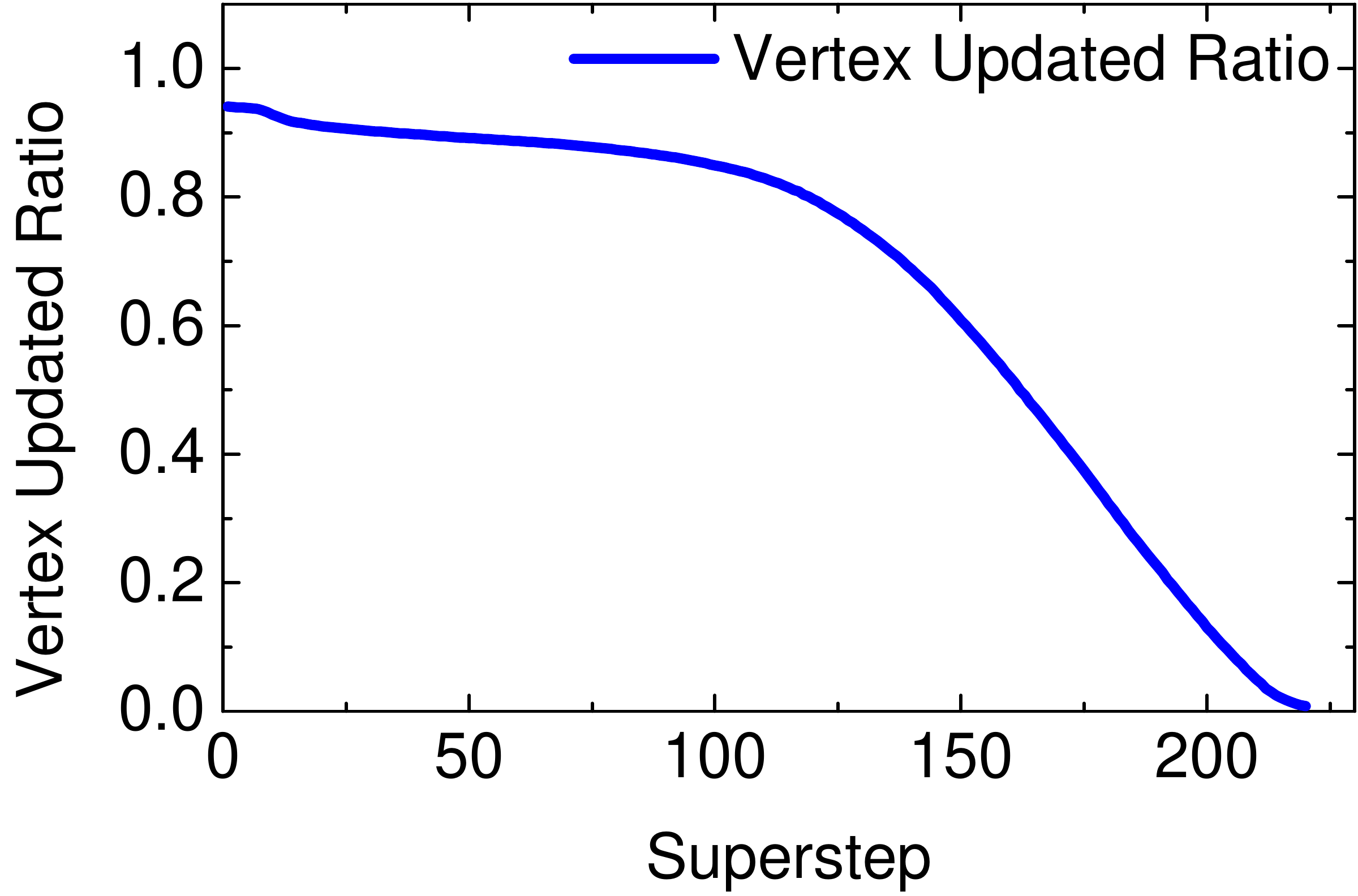}
    \subcaption{(a) Vertex Update Ratio.}
    \end{center}
    \end{minipage}
    \centering
    \begin{minipage}[t]{\minipagewidth}
    \begin{center}
    \includegraphics[width=\figurewidthFour]{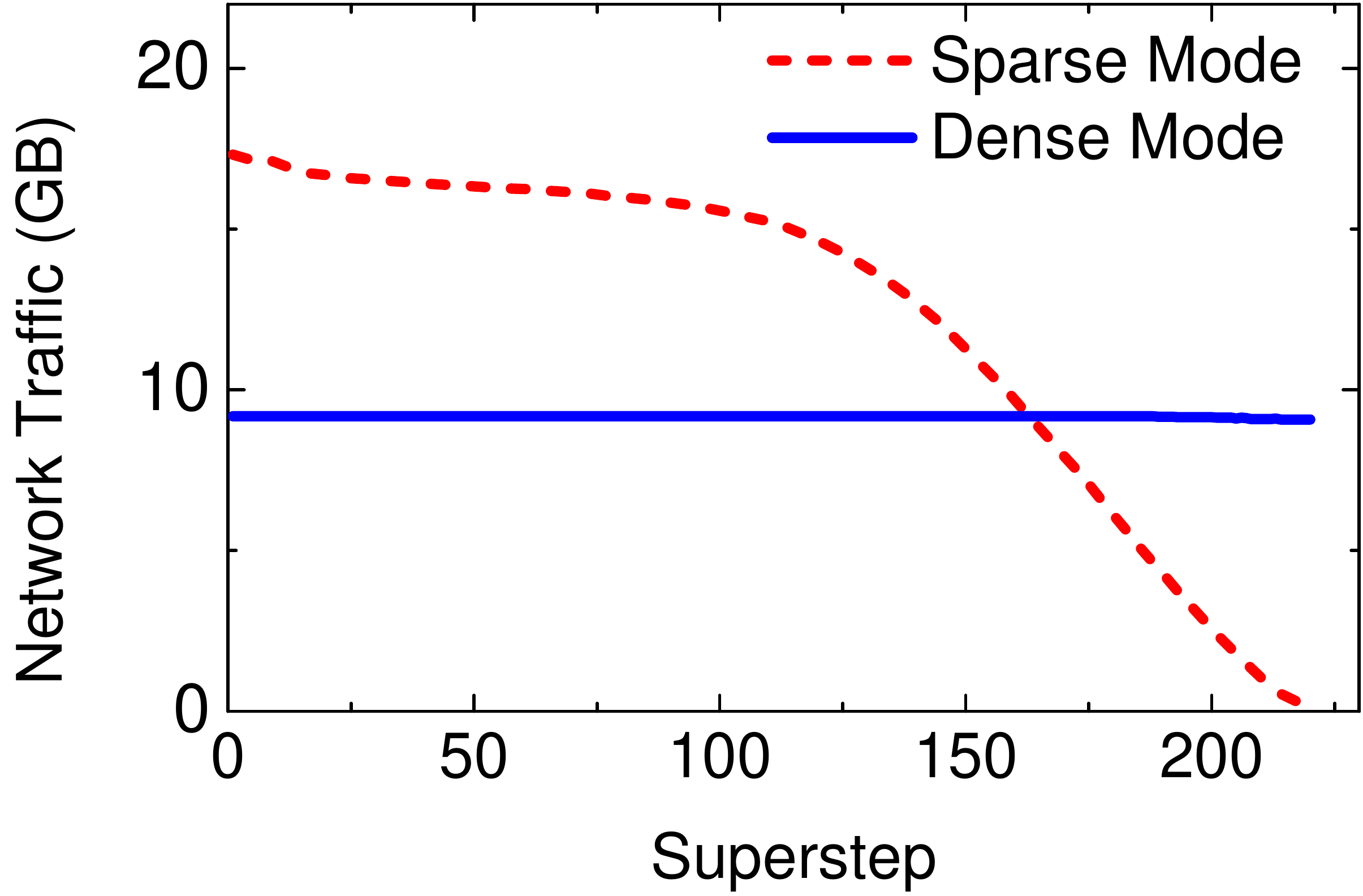}
    \subcaption{(b)  Network Traffic (Sparse/Desnse Mode).}
    \end{center}
    \end{minipage}
    \centering
    \begin{minipage}[t]{\minipagewidth}
    \begin{center}
    \includegraphics[width=\figurewidthFour]{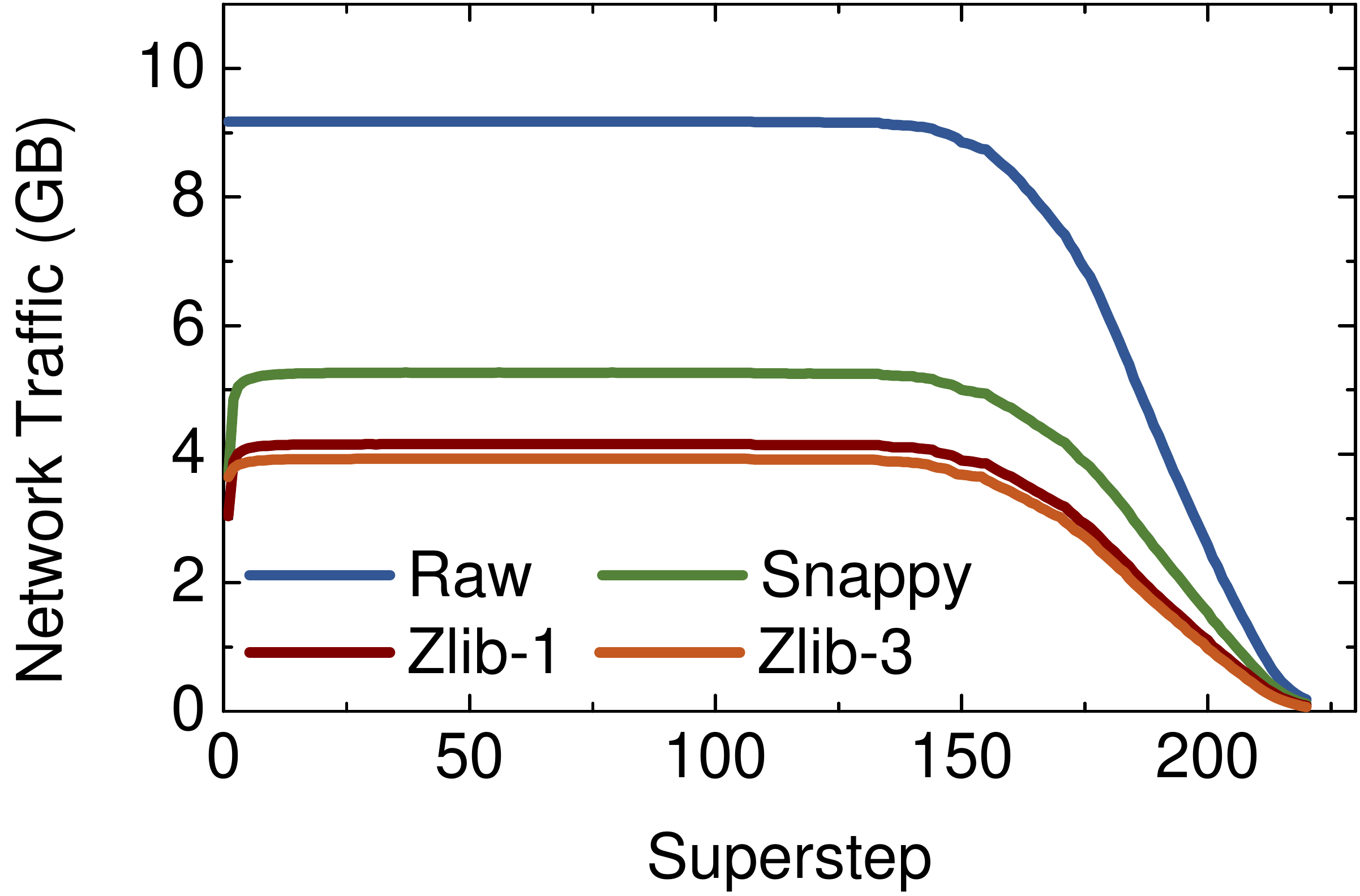}
    \subcaption{(c)  Network Traffic (Hybrid Mode).}
    \end{center}
    \end{minipage}
    \centering
    \begin{minipage}[t]{\minipagewidth}
    \begin{center}
    \includegraphics[width=\figurewidthFour]{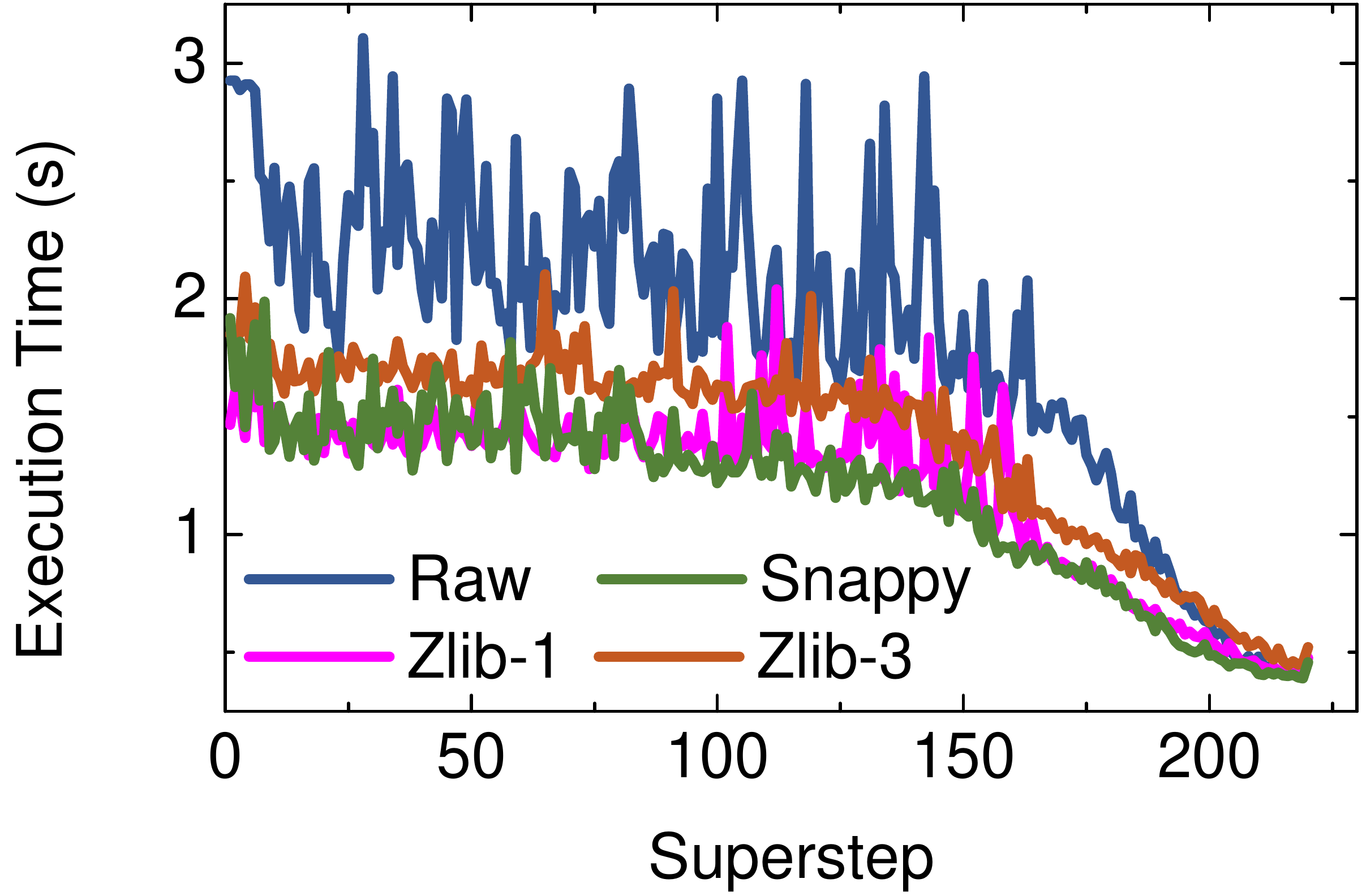}
    \subcaption{(d)  Execution Time (Hybrid Mode).}
    \end{center}
    \end{minipage}
    \centering
    \caption{Network traffic of PageRank on UK-2007 in a 9-node cluster.}
\label{Fig: hybridnetwork}
\end{figure}

We design a hybrid communication mode to save network bandwidth. Specifically, GraphH firstly uses a dense array to store all updated vertex values. Before the broadcasting phase, GraphH checks its sparsity ratio \footnote{Sparsity ratio is measured by  the unchanged vertex number over the total vertex number of a tile.}. If the sparsity ratio is higher than a given threshold (in this paper, this threshold is set to 0.8), GraphH converts it into a sparse array, which only stores non-zero values and their indices. Take PageRank as an example, as shown in Figure \ref{Fig: hybridnetwork} (c), at the beginning of the program, GraphH broadcasts messages under the dense node. At the end of the program, GraphH would switch to the sparse communication mode to avoid sending zeros. 

Message compressing can further improve communication performance as in \cite{sun2016timed}. Figure \ref{Fig: hybridnetwork} (c) shows that  snappy, zlib-1 and zlib-3 could reduce network traffic by a factor of 1.7,  2.3 and 2.3, respectively.
Figure \ref{Fig: hybridnetwork} (d) shows that reduced network traffic can lead to improved graph processing performance. In the first 50 supersteps, GraphH roughly takes 2.32s per superstep without compression. When using snappy, zlib-1 and zlib-3, GraphH averagely takes 1.73s, 1.56s and 1.5s per superstep. 
While zlib could save more network traffic, the additional decompressing overhead makes it consume more execution time than snappy. The default network traffic compressor of GraphH is snappy.

\setlength{\minipagewidth}{0.238\textwidth}
\setlength{\figurewidthFour}{\minipagewidth}
\begin{figure} 
    \centering
    \begin{minipage}[t]{\minipagewidth}
    \begin{center}
    \includegraphics[width=\figurewidthFour]{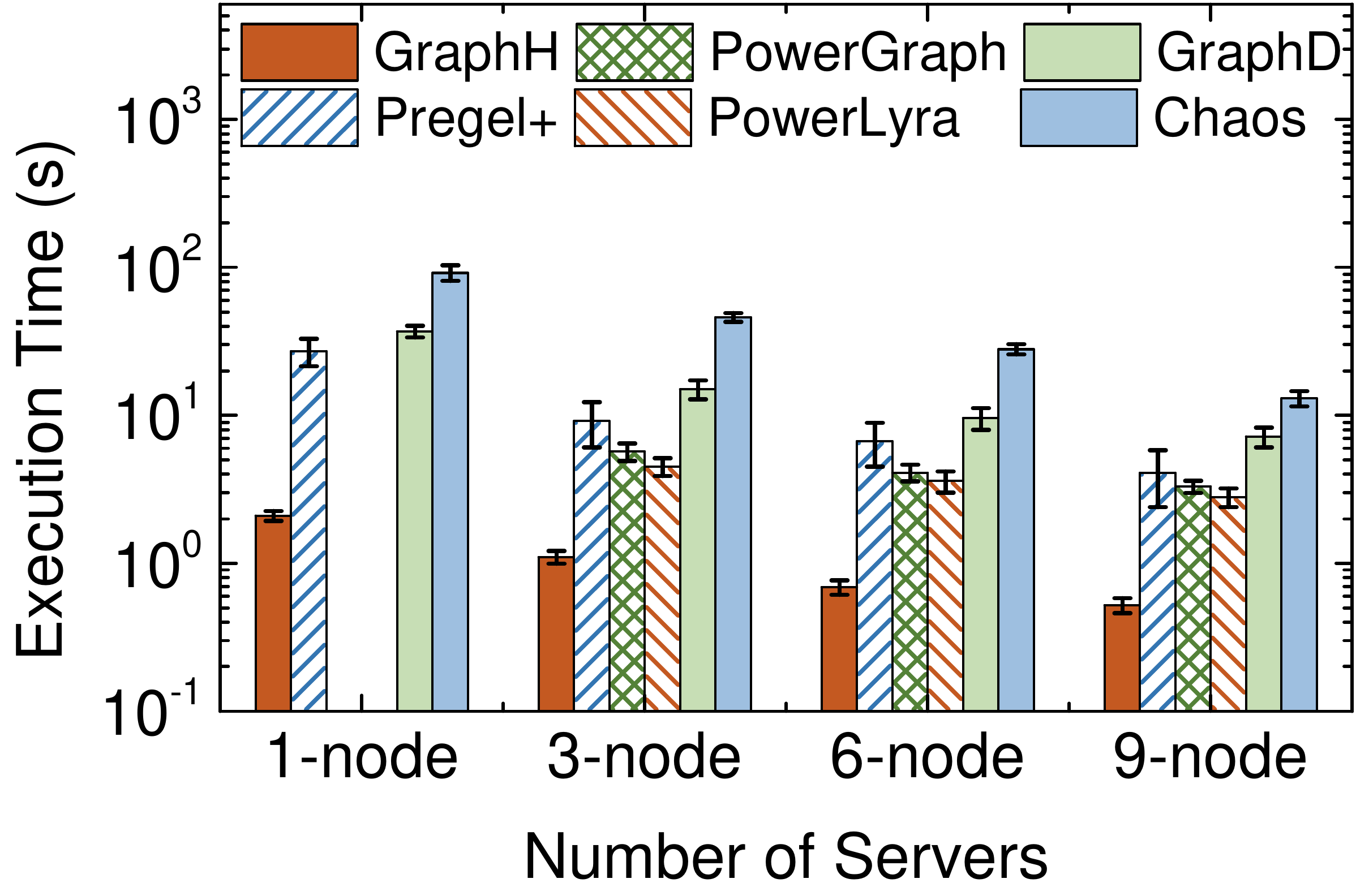}
    \subcaption{(a) Twitter-2010}
    \end{center}
    \end{minipage}
    \centering
    \begin{minipage}[t]{\minipagewidth}
    \begin{center}
    \includegraphics[width=\figurewidthFour]{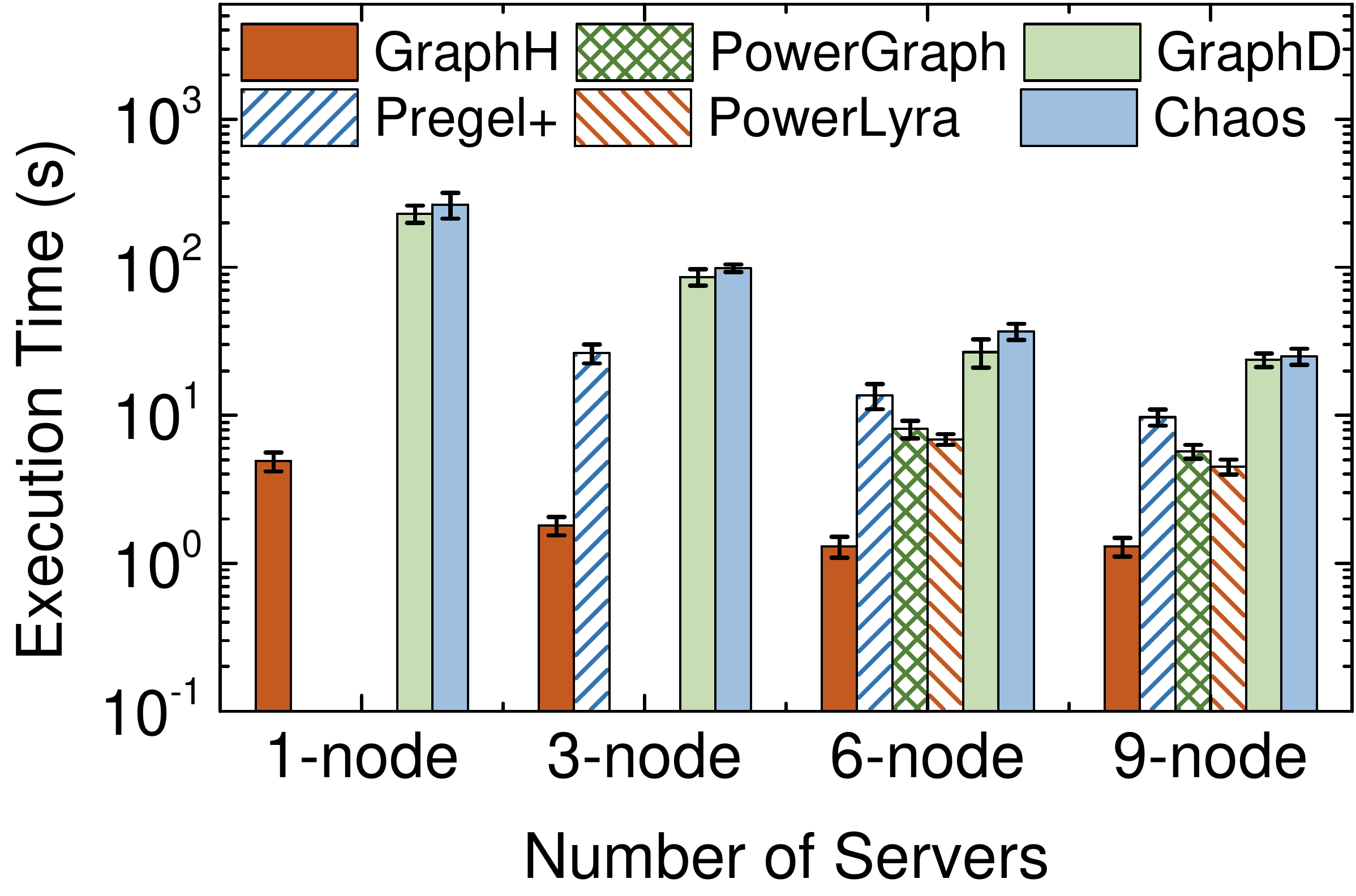}
    \subcaption{(b) UK-2007}
    \end{center}
    \end{minipage}
    \centering
    \begin{minipage}[t]{\minipagewidth}
    \begin{center}
    \includegraphics[width=\figurewidthFour]{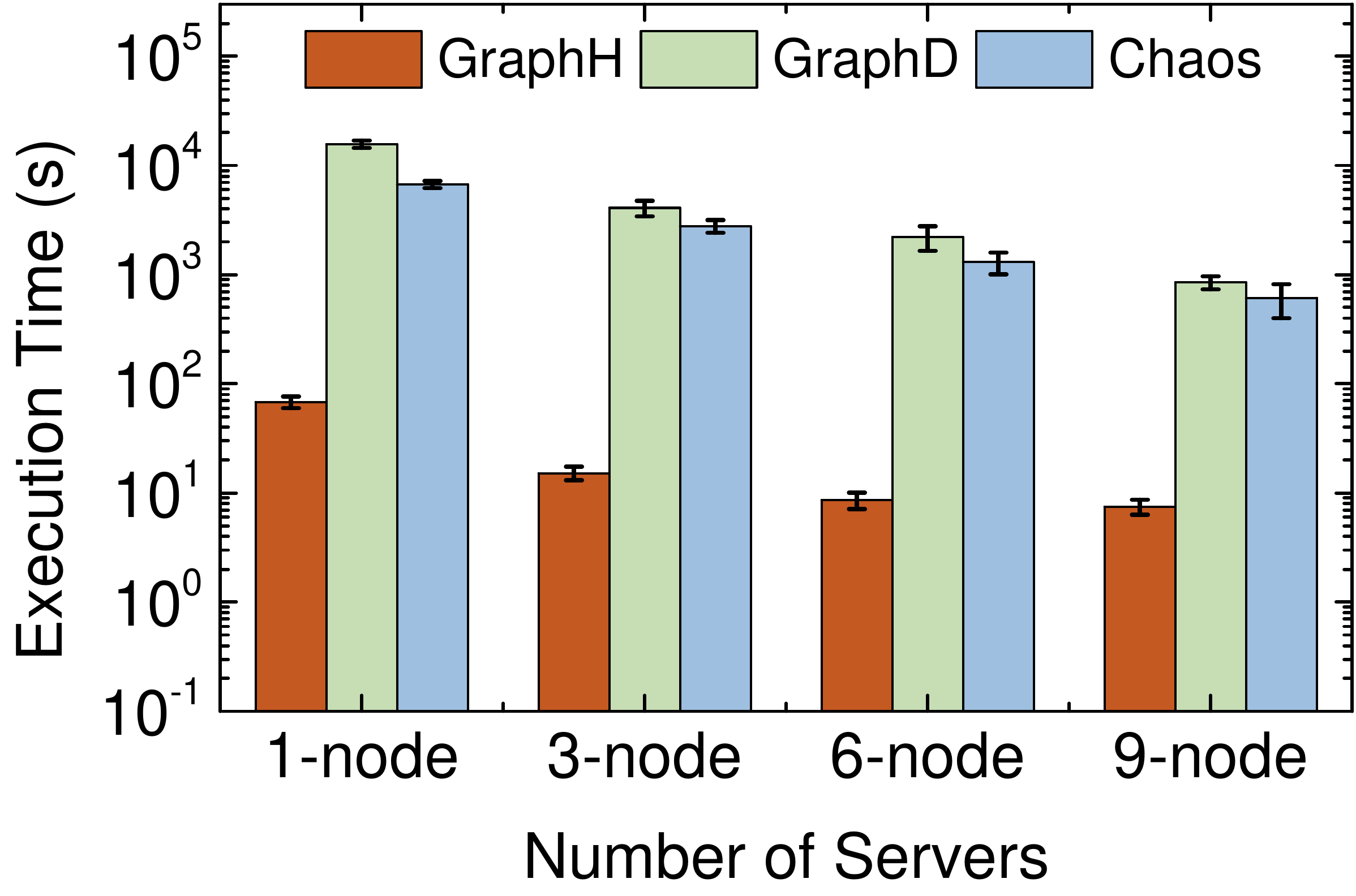}
    \subcaption{(c) UK-2014}
    \end{center}
    \end{minipage}
    \centering
    \begin{minipage}[t]{\minipagewidth}
    \begin{center}
    \includegraphics[width=\figurewidthFour]{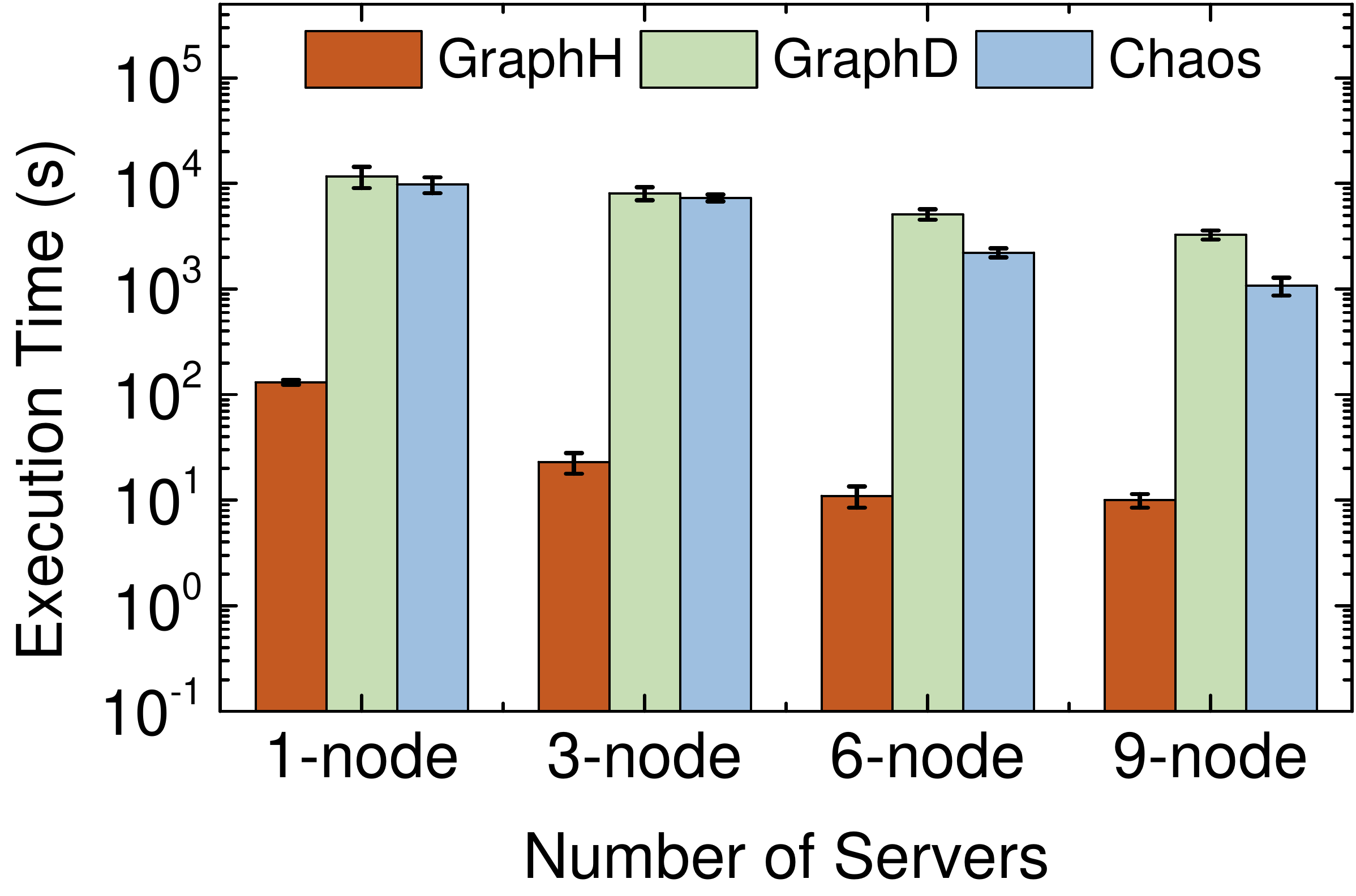}
    \subcaption{(d) EU-2015}
    \end{center}
    \end{minipage}
    \centering
    \caption{The average execution time per superstep  to run PageRank.}
\label{Fig: PageRankResult}
\end{figure}

\section{Evaluation Results}

In this section, we evaluate GraphH's performance using a testbed with two applications (PageRank, SSSP) and four graphs (Twitter-2010, UK-2007, UK-2014 and EU-2015). The hardware and software configurations are same with the testbed shown in Figure \ref{Fig: MemUsage_and_Time}. We use the average execution time per superstep as the performance metric. 
For each experiment, we run 21 supersteps, and calculate the average execution time without the first superstep, since distributed graph processing systems usually load the input graph to memory during this superstep. We do not compare the graph loading time, since existing graph processing systems do not have unique storage system requirements. For example, Pregel+ and GraphD only support HDFS. In Chaos, 
we must manually distributed the graph partitions to multiple servers. 

\subsection{Performance of PageRank}

Figure \ref{Fig: PageRankResult} shows that GraphH can achieve much higher performance than Pregel+, PowerGraph, PowerLyra, GraphD and Chaos when running PageRank on all four input graphs. Moreover, we observe that GraphH's memory management strategy is efficient, since it can even process big graphs like UK-2014 and EU-2015 in a single node.  

When running PageRank on Twitter-2010 with 9 servers, GraphH could outperform Pregel+, PowerGraph, PowerLyra, GraphD and Chaos by 7.8x, 6.3x, 5.3x, 13x and 25x, respectively. The corresponding speedup ratios for UK-2007 are 7.5, 4.3, 3.5, 18 and 19. The performance gain comes from GraphH's reduced communication overhead.

As shown in Figure \ref{Fig: PageRankResult} (c) (d),  GraphH can efficiently run PageRank on big graphs like UK-2014 and EU-2015 in a small cluster or even a single machine with limited memory.  More specifically, on a single node, GraphH only takes 68s and 131s to run a superstep of PageRank on UK-2014 and EU-2015, respectively. When having a cluster with 9 server, GraphH only needs 7.5s and 10s on average per superstep, and could roughly outperforms GraphD and Chaps by 320 and 110, respectively.
The performance gain comes from the reduced disk I/O overhead of GraphH. 

\setlength{\minipagewidth}{0.238\textwidth}
\setlength{\figurewidthFour}{\minipagewidth}
\begin{figure} 
    \centering
    \begin{minipage}[t]{\minipagewidth}
    \begin{center}
    \includegraphics[width=\figurewidthFour]{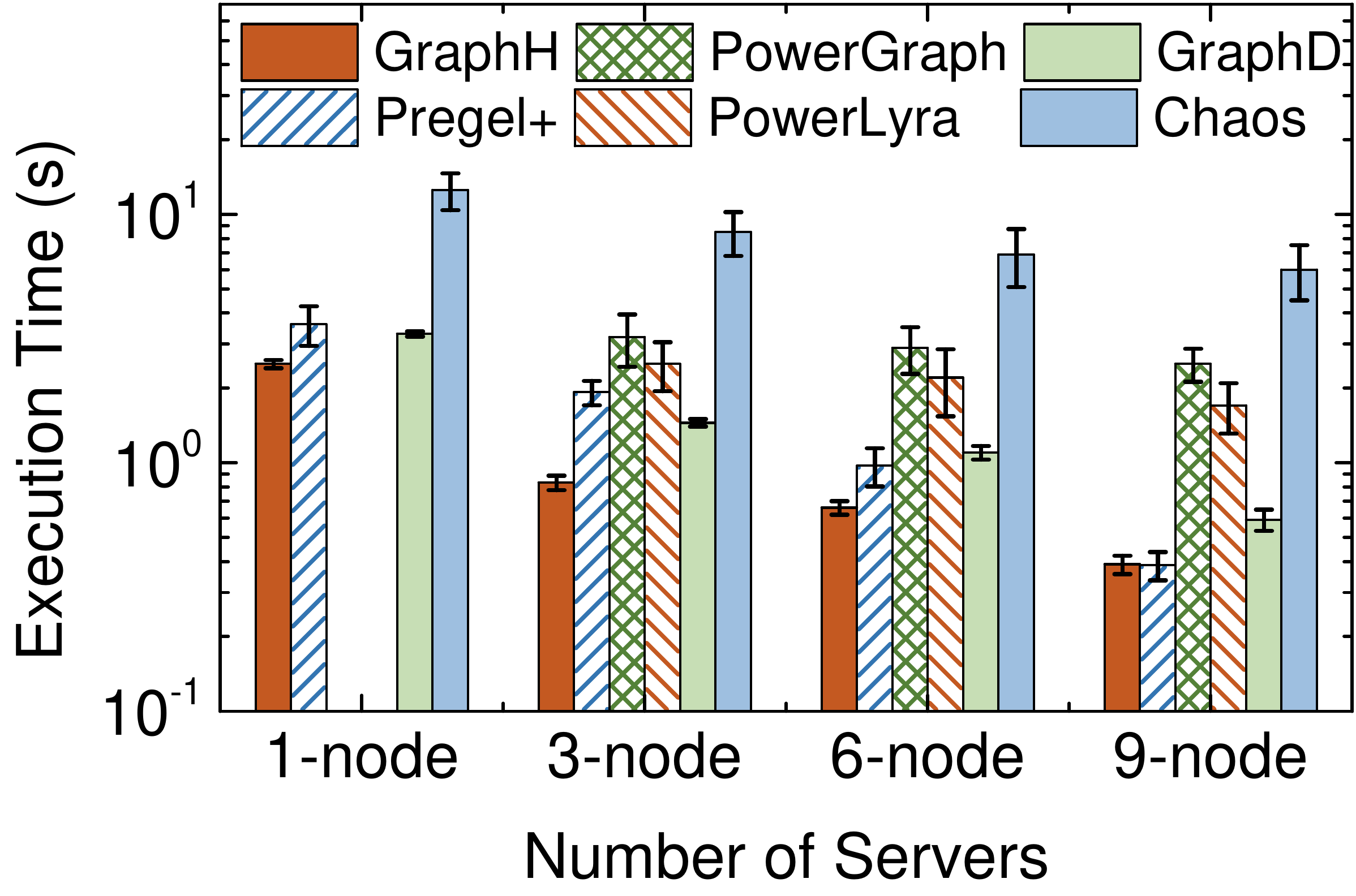}
    \subcaption{(a) Twitter-2010}
    \end{center}
    \end{minipage}
    \centering
    \begin{minipage}[t]{\minipagewidth}
    \begin{center}
    \includegraphics[width=\figurewidthFour]{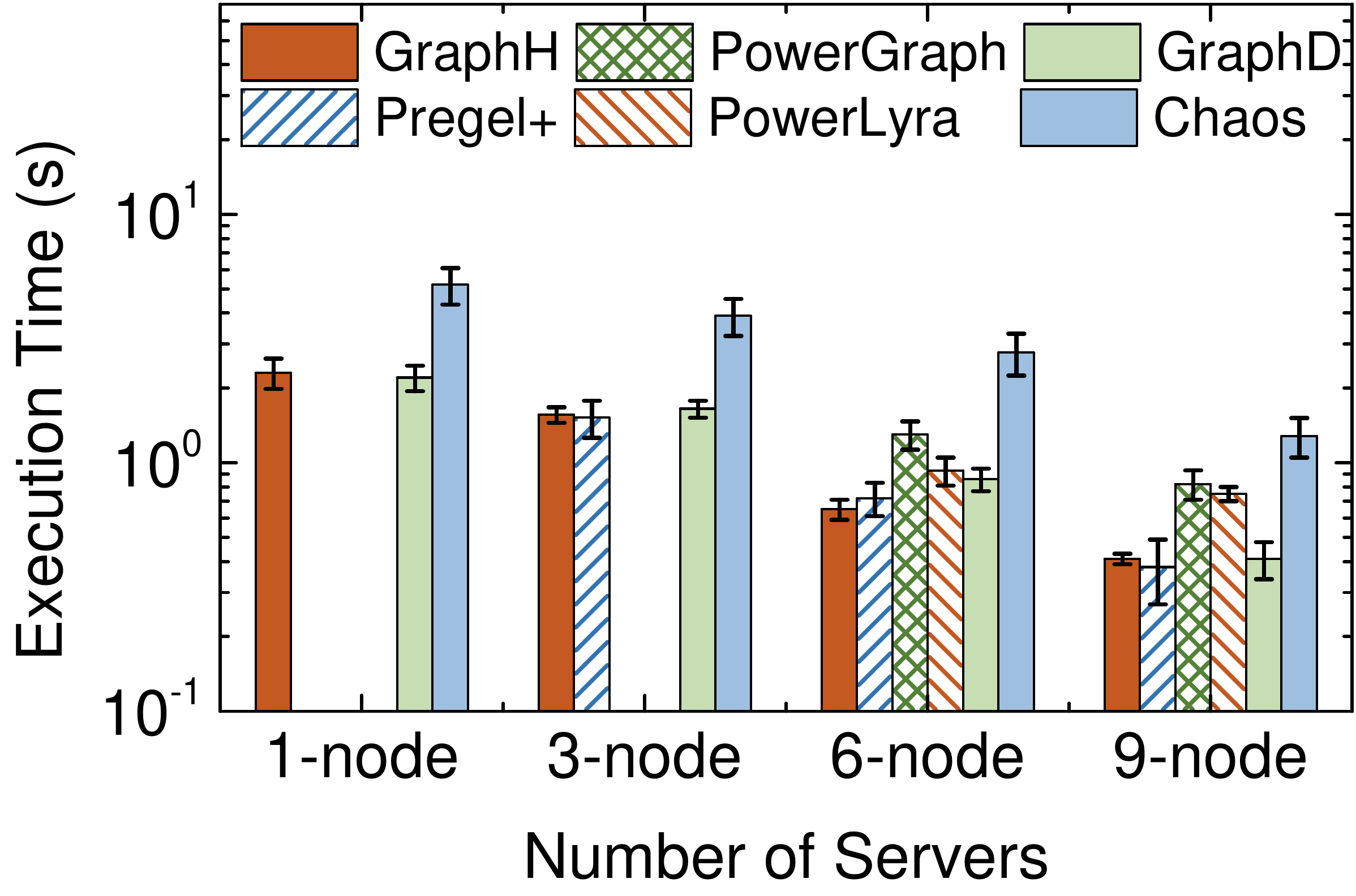}
    \subcaption{(b) UK-2007}
    \end{center}
    \end{minipage}
    \centering
    \begin{minipage}[t]{\minipagewidth}
    \begin{center}
    \includegraphics[width=\figurewidthFour]{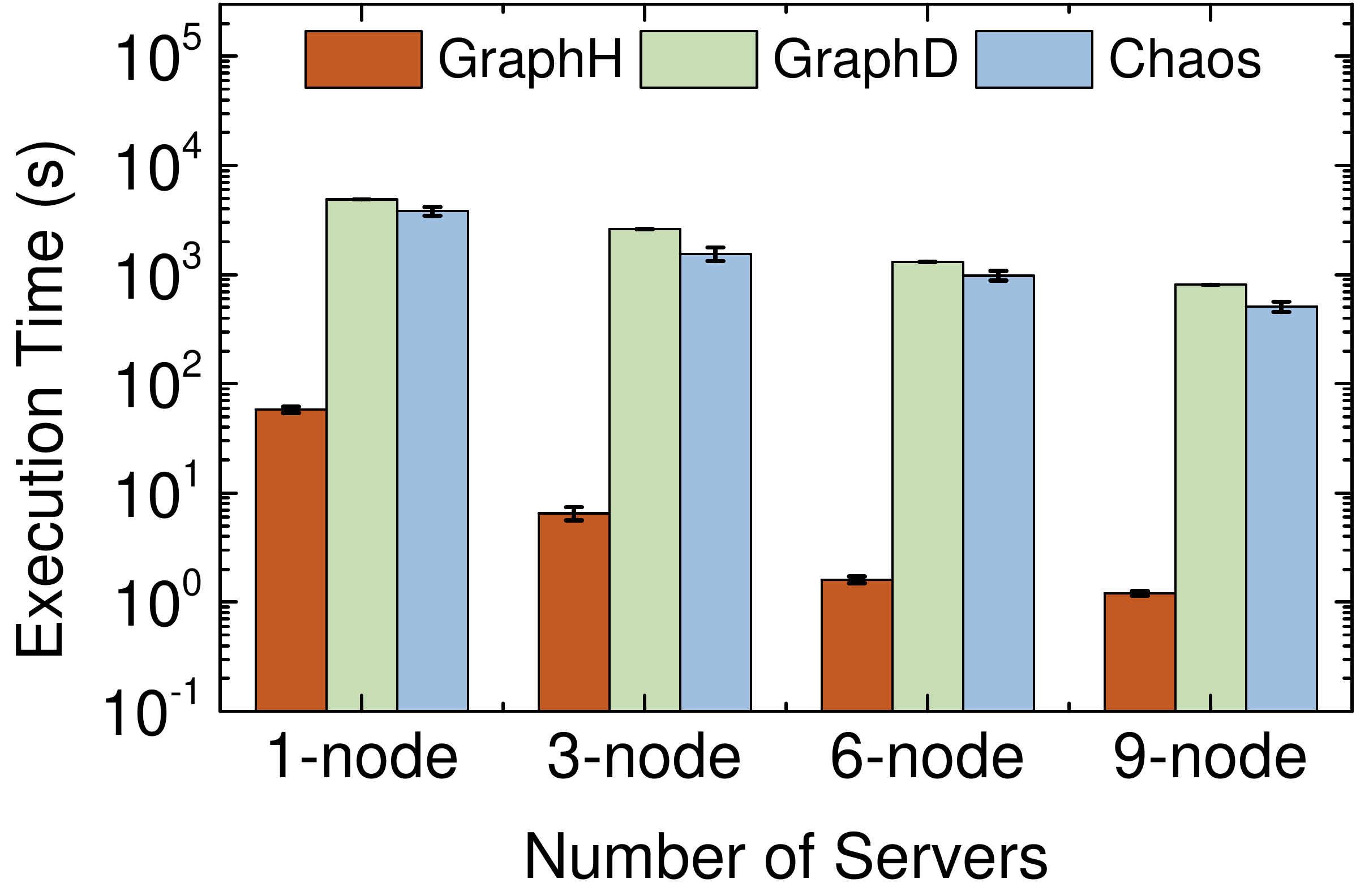}
    \subcaption{(c) UK-2014}
    \end{center}
    \end{minipage}
    \centering
    \begin{minipage}[t]{\minipagewidth}
    \begin{center}
    \includegraphics[width=\figurewidthFour]{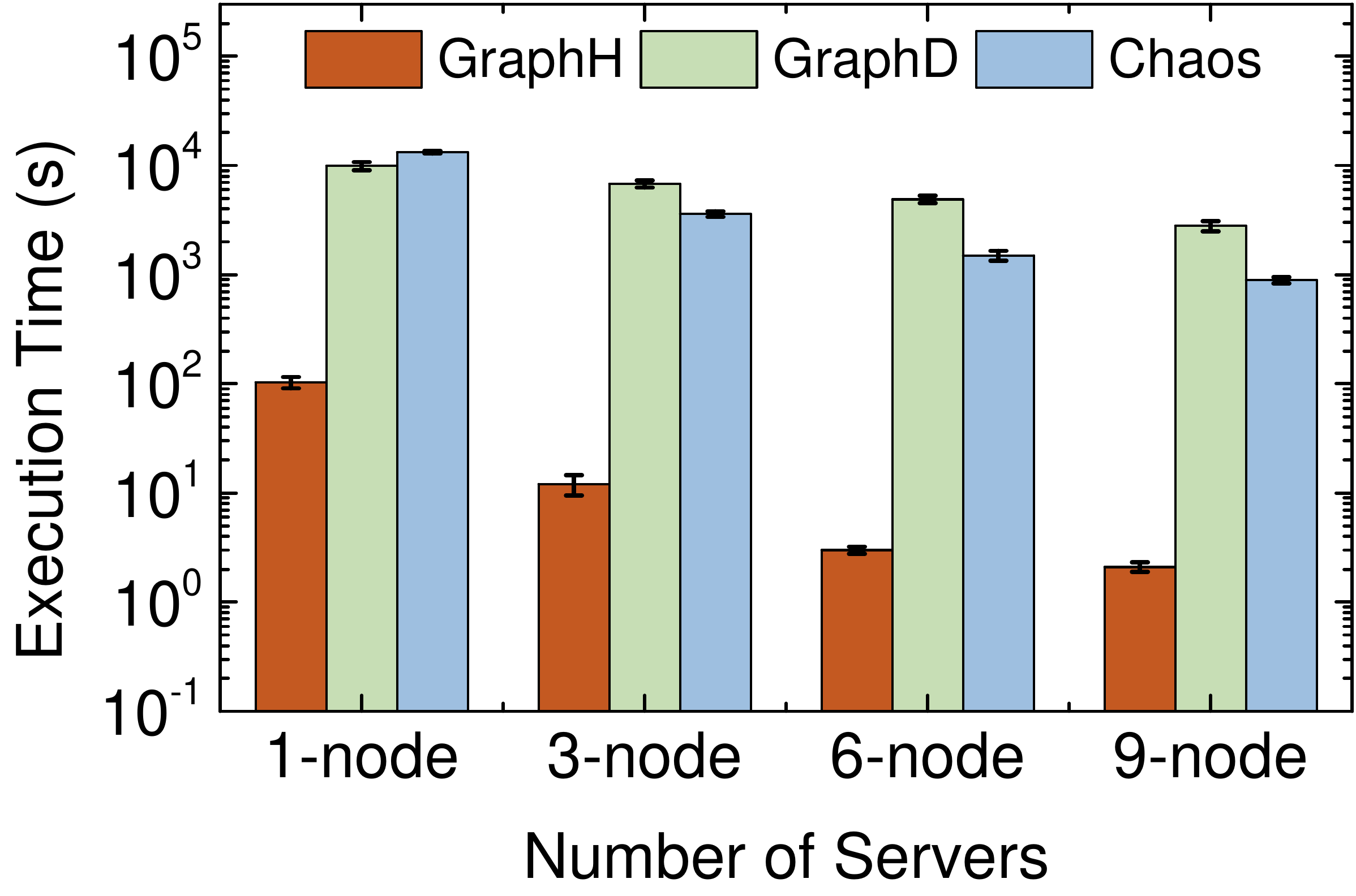}
    \subcaption{(d) EU-2015}
    \end{center}
    \end{minipage}
    \centering
    \caption{The average execution time per superstep  to run SSSP.}
\label{Fig: SSSPResult}
\end{figure}

\subsection{Performance of SSSP}

Figure \ref{Fig: SSSPResult} shows that GraphH also works well with SSSP. When running SSSP on Twitter-2010 and UK-2007 with 9 servers, GraphH has a similar performance with Pregel+, and both of them roughly takes 0.4s to run a  superstep. The reason is that the communication overhead is not significant for SSSP, since only a small partition of vertices may update their values. In addition, GraphH outperforms PowerGraph and PowerLyra by up to 2 for running SSSP on Twitter-2010 and UK-2007. 
From Figure \ref{Fig: SSSPResult} (c) (d),  we can see that GraphH's memory management is efficient when running SSSP on big graphs. Specifically,  GraphH roughly takes 1.3s and 2.1s to run a superstep of SSSP on UK-2014 and EU-2015 with 9 servers. Since GraphD and Chaos have high disk I/O overhead, GraphH could outperform them by at least 350x.

\section{Conclusion}
 
In this paper, we tackle the challenge of big graph analytics in small clusters. Existing in-memory systems need a huge amount of resources to handle big graphs, and out-of-core systems have poor performance due to high disk I/O overhead.
We propose a new distributed graph processing system named GraphH, which does not require to store all data in memory, but it maximizes the amount of in-memory data.
GraphH partitions the input graph into tiles, and  makes each worker process a tile in memory  at a time to reduce memory footprint. We design an edge cache system to reduce disk I/O overhead, and propose a hybrid  approach to reduce  communication overhead. As a result, GraphH can efficiently process big graphs with limited memory.
Extensive evaluations show that GraphH could outperform existing in-memory systems by up to 7.8x when processing generic graphs, and outperform existing out-of-core systems by more than 100x when processing big graphs.

\section*{Acknowledgment}
\noindent
This work is supported by the Data Science and AI Research Center of the Nanyang Technological University, Singapore.


\balance

\bibliographystyle{IEEEtran}
\bibliography{main}


\end{document}